# Dynamic Slack-Aware Clocking for Near-Threshold Tensor Processing Units (TPUs)

MUHAMMAD USMAN NADEEM, Utah State University, USA
SANGHAMITRA ROY, Utah State University, USA
KOUSHIK CHAKRABORTY, Utah State University, USA

**Operating Tensor Processing Units (TPUs) in the near-threshold computing (NTC) region significantly reduces energy consumption but introduces high delay sensitivity to process variation and data activity. Conventional designs typically rely on a conservative, fixed global clock to ensure safety, which leaves large portions of timing margin unexploited as most operations finish well before the clock edge. We propose Dynamic Slack-Aware Clocking (DSAC), a proactive framework that replaces worst-case timing with operation-specific adjustments. DSAC employs lightweight Hamming-Distance, Most-Significant-Bit, and Hybrid predictors to estimate the delay sensitivity of individual multiply-accumulate (MAC) operations and classify them into three timing tiers. These tiers are enforced locally via dummy-hold cycles under a fixed global reference clock, enabling fine-grained timing adaptation without global clock retuning or frequency scaling. A closed-loop feedback controller monitors timing violations and updates tier thresholds at runtime to maintain resilience. Experiments on quantized DNN benchmarks demonstrate that the MSB predictor maintains high inference accuracy, with an average loss of only 1% even at aggressive performance points. Furthermore, DSAC achieves up to 1.55× better energy efficiency at 2.15× frequency scaling compared to a baseline TPU, while incurring an area overhead as low as 13%.**

## 1 Introduction

THE unprecedented growth of deep neural network (DNN) workloads has intensified the need for accelerators that deliver high performance with low energy consumption. TPUs [1], equipped with large systolic arrays [2] of multiply–accumulate (MAC) units, have become the cornerstone of modern AI infrastructure, enabling orders-of-magnitude speedup over CPUs and GPUs [3]. However, this massive parallelism often results in high power consumption that is increasingly unsustainable for continuous inference workloads in datacenters.

Operating TPUs in the near-threshold computing (NTC) region [4] significantly reduces energy use [5, 6], but also introduces extreme delay sensitivity to process variation and data activity. Consequently, designers must choose between

Authors' Contact Information: Muhammad Usman Nadeem, Utah State University, Electrical and Computer Engineering, Logan, Utah, USA, usman.nadeem@usu.edu; Sanghamitra Roy, Utah State University, Electrical and Computer Engineering, Logan, Utah, USA, sanghamitra.roy@usu.edu; Koushik Chakraborty, Utah State University, Electrical and Computer Engineering, Logan, Utah, USA, koushik.chakraborty@usu.edu.

This is the author-accepted version of a paper accepted for publication in *ACM Transactions on Design Automation of Electronic Systems (TODAES)*. It is not the final ACM-formatted Version of Record. The Version of Record is available at https://doi.org/10.1145/3837085.

aggressive frequency scaling that risks frequent timing errors or conservative fixed clocks that leave substantial timing slack—the margin between circuit delay and the clock period—underutilized. This fundamental inefficiency necessitates a finer-grained control mechanism capable of dynamically adapting timing at the level of individual MAC operations.

Prior efforts have explored various ways to handle timing violations in low-voltage environments. Razor [7] detects late signals but requires costly pipeline replays, which are unsuitable for the rigid dataflow of deep systolic arrays. TE-Drop [8] corrects errors by bypassing downstream MACs, at the cost of accuracy loss when errors accumulate. More recent designs such as GreenTPU [5] and EFFORT [6] improve resilience and efficiency through predictive boosting and same-cycle correction. While effective, these schemes remain largely reactive, addressing errors only after they occur.

In this work, we adopt a proactive approach that predicts and adapts to timing variation before errors arise. We observe that MAC delay exhibits strong data dependence, driven by both operand significance and input switching activity. As shown by our slack profiling in Section 2.1, we leverage this behavior to propose *Dynamic Slack-Aware Clocking (DSAC)*, a family of lightweight prediction-based dynamic clocking techniques:

- a **Hamming-Distance (HD) predictor**, which measures the number of bit transitions between consecutive activation inputs to capture dynamic switching activity;
- a **Most-Significant-Bit (MSB) predictor**, which identifies operand significance to approximate static path delay; and
- a **Hybrid predictor**, which combines both HD and MSB characteristics to jointly model cycle-to-cycle delay variation.

These predictors guide a three-tier adaptive clocking framework that adjusts the effective timing budget of each MAC operation—providing fast execution for non-critical cases and extended execution windows for delay-critical ones. Key contributions include:

- We propose a fine-grained dynamic clocking scheme for near-threshold TPUs that replaces fixed worst-case timing with operation-specific adjustments to exploit unused timing margin.
- We develop three complementary prediction models—HD, MSB, and Hybrid—and analyze their impact on timing slack utilization and error resilience in the NTC region.
- We propose a tier-based timing control architecture that maps predicted criticality to three pre-characterized timing tiers via a compact lookup table (LUT) and locally enforced dummy-hold cycles, enabling fine-grained adaptation without the overhead of global clock changes or DVFS.
- We implement the proposed predictors in a cycle-accurate systolic array simulator integrated with a statistical timing analysis (STA) backend to model per-MAC delay behavior under process and data variation.
- We demonstrate that the MSB and HD predictors maintain the highest inference accuracy across the frequency range, while the Hybrid predictor consistently reduces underutilized timing margin with an average accuracy loss of less than 2%.

By anticipating timing-critical cycles and adapting the local timing budget accordingly, the proposed framework delivers high performance and energy efficiency in near-threshold TPUs—effectively bridging the gap between reactive error correction and proactive timing management.

## 2 Motivation

In this section, we highlight the opportunity to reduce unused timing margin—the slack left by early-finishing MAC operations—to improve both energy efficiency and performance of NTC TPUs. Section 2.1 reviews the TPU systolic array architecture and existing error-handling schemes. Section 2.2 outlines our measurement methodology. Section 2.3 presents empirical evidence from large-scale timing analysis, and Section 2.4 discusses the insights that motivate our proposed *DSAC* framework.

## 2.1 Background

Matrix multiplication dominates the computation in DNNs and is accelerated in TPUs using large systolic arrays of MAC units [9]. The weight matrices are preloaded into the array, activations stream horizontally, and partial sums propagate vertically, enabling massive throughput compared to CPUs and GPUs.

Operating TPUs in the low-voltage or near-threshold regime dramatically reduces energy consumption but also increases timing variability due to slower transistors and data-dependent delay. .To ensure correctness, designers typically choose a single global clock long enough for the slowest possible MAC [10]. This fixed worst-case clock guarantees safe execution but leaves the majority of operations finishing early, creating substantial wasted slack.

Reactive resilience schemes such as Razor [7], TE-Drop [8], and EFFORT [6] attempt to recover from rare violations when the clock is tightened. However, these methods remain fundamentally reactive and input-agnostic—addressing errors only after they occur while failing to exploit the abundant positive slack present in most cycles. This result is a persistent underutilization of the timing margin, wasting both execution time and potential energy savings.

## 2.2 Methodology

To quantify the variation of timing slack across operations, we developed a cross-layer profiling framework that integrates circuit-level delay analysis with architectural-level simulation. At the circuit level, an in-house STA engine, implemented in C++, was utilized to extract path delays of synthesized MAC units under specific operand switching patterns. The engine accounts for process variation by perturbing a random subset of gates according to established NTC models [8, 6]. This STA engine was integrated into a cycle-accurate systolic array simulator, enabling the measurement of per-cycle delay and slack during realistic workloads

For our experiments, we executed 500 randomly generated 8×8 activation tiles. Weights were preloaded into the systolic array, and activations streamed through it cycle by cycle. For each MAC operation, the STA reported the actual path delay $D$, compared against the nominal clock period $T_{clk}$. Slack was recorded as slack=$T_{clk}$–$D$. In total, this produced over 500,000 slack samples. The resulting normalized distribution is shown in Figure 1.

## 2.3 Experimental Evidence

Figure 1 illustrates the distribution of the Required Cycle Fraction (RCF), defined as the ratio of an operation's delay $D_t$ to the nominal clock period $T_{\text{clk}}$:

$$\text{RCF} = \frac{D_t}{T_{\text{clk}}}.$$

The histogram exhibits four dominant delay modes with peaks near 20%, 45%, 65%, and 120%, indicating a strongly multi-modal delay landscape. Importantly, around 80% of MAC operations fall below the 1.0 boundary, completing within a single cycle with substantial positive slack. The dominant mass near RCF ≈65% further shows that most operations finish with roughly 35% of the cycle remaining. In contrast, only a small tail (15–20%) exceeds the deadline, yet these rare outliers determine the global clock period in conventional fixed-frequency designs. Consequently, worst-case clocking systematically converts common-case positive slack into wasted timing margin. This quantitative imbalance motivates DSAC, which reclaims the available margin by replacing global conservatism with operation-specific timing tiers.

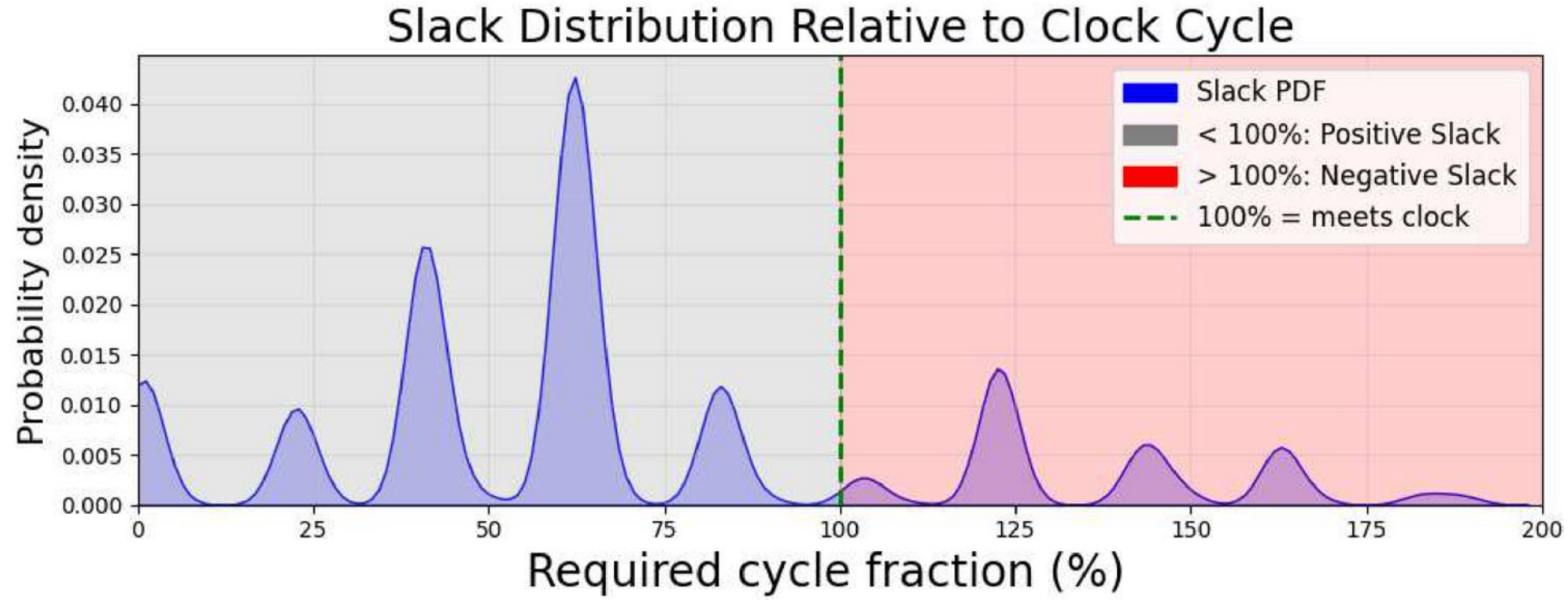

**Fig. 1.** *Distribution of normalized slack across 500,000 MAC operations. Over 80% of operations complete within 65% of the clock cycle, leaving roughly one-third of the timing margin unexploited under conventional fixed-clock design.*

### 2.4 Significance and Insights

The measured distribution exposes a critical inefficiency in near-threshold TPUs: vast timing slack remains unexploited across most operations. Reducing this waste offers a powerful opportunity to enhance both performance and energy efficiency without compromising correctness.

Further analysis shows that delay variation correlates strongly with operand-dependent characteristics:

- **Hamming-distance effect:** Cycles with a higher number of bit transitions between consecutive activations exhibit greater switching activity, increasing dynamic delay.
- **MSB significance effect:** Operands with more high-order bits set activate deeper logic paths in the multiplier, leading to longer propagation delays.

These two effects jointly determine how much of the timing budget a MAC consumes in each cycle. Recognizing them enables the prediction of “fast” and “slow” cycles ahead of time.

Based on this observation, we propose DSAC, which proactively tailors the effective clock period of each MAC using lightweight predictors:

(1) a **Hamming-Distance (HD) predictor** that models temporal switching activity,
(2) an **MSB-based predictor** that captures static operand significance, and
(3) a **Hybrid predictor** that fuses both effects to maximize slack utilization.

By dynamically allocating execution time only where it is required, DSAC minimizes wasted slack and recovers the performance and energy lost to conventional worst-case clocking—enabling efficient and reliable near-threshold TPU operation.

## 3 Design

This section describes the architectural framework of the proposed design, which builds upon the timing-variation analysis and motivation discussed in Section 2. It first presents a high-level overview of the approach and its operational flow, then details the predictor architecture, tier mapping strategy, and local timing-control mechanisms that collectively enable energy-efficient, error-resilient operation in near-threshold TPUs.

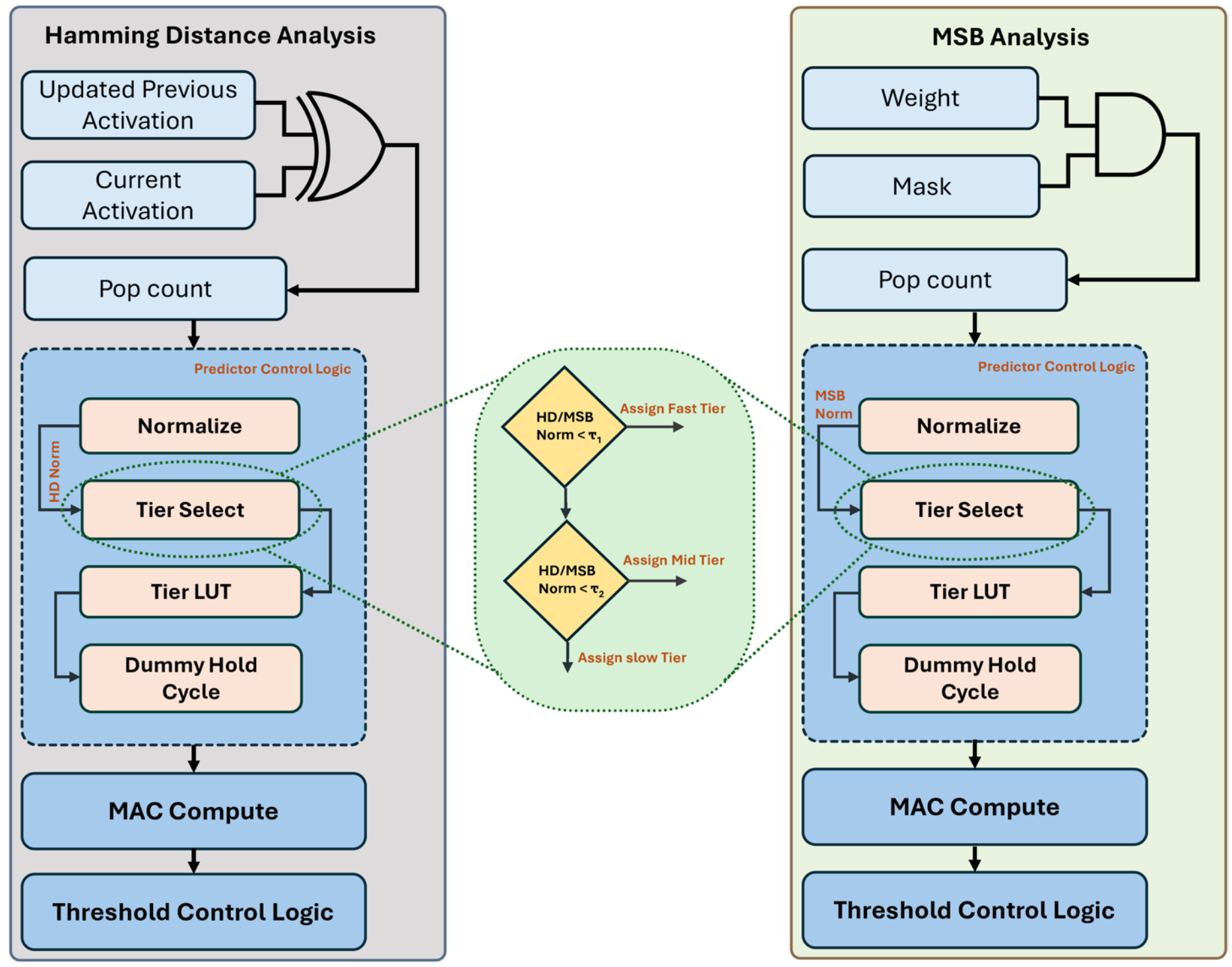

**Fig. 2.** *Hardware implementation of the HD and MSB predictors. Using only XOR and popcount, HD estimates activation switching and MSB estimates operand magnitude (upper-half bit density) to select one of three timing tiers, with MSB incurring as little as 13% area overhead.*

### 3.1 Overview

To enhance timing-error resilience while maintaining high energy efficiency, we propose a Dynamic Clocking framework for near-threshold systolic arrays. Unlike fixed-frequency designs that enforce a single worst-case clock, DSAC exploits the predictable dataflow of TPUs to assign a data-dependent timing budget to each MAC operation. Lightweight predictors evaluate per-operation delay sensitivity based on operand features and classify operations into one of three pre-characterized timing tiers.

We utilize a single fixed fast reference clock with period $T_{\text{fast}}$, generated by the standard SoC clock generator phase-locked loop (PLL) [11]. This PLL is not part of DSAC and remains static during inference; DSAC only gates or holds local pipeline registers to implement hold cycles for mid and slow tiers without modifying the global clock. Each processing element (PE) realizes an operation-specific effective timing budget $T_{\text{eff}}$ by locally inserting dummy-hold cycles (stalling for a few $T_{\text{fast}}$ cycles) for Mid/Slow operations. The fast-clock frequency remains fixed during execution, and DSAC avoids the overhead of global DVFS [12] or per-cycle clock retuning.

STA provides the actual delay $D_t$ for each MAC. Slack is computed as $T_{\text{eff}}-D_t$, where $T_{\text{eff}}$ is the effective timing budget selected from the tier LUT. Because most MACs operate well below the worst-case delay, enforcing a single long global clock wastes slack and degrades throughput. Our framework instead leverages intra-array timing heterogeneity, allowing only the rare “hard” operations to incur extended execution through local dummy holds rather than global DVFS

**Hamming Distance Predictor**

Prev = 11110000
Curr = 00001111
XOR = 11111111 ⇒ many toggles (slow)

Prev = 11110000
Curr = 11110001
XOR = 00000001 ⇒ few toggles (fast)

**MSB Predictor**

Operand = 1111 0000 (many upper-half bits set)
Large magnitude ⇒ deeper carry path ⇒ slow

Operand = 0011 0000 (only bits 5:4 high in upper half)
Fewer upper-half bits ⇒ shallower path ⇒ fast

**Hybrid Predictor**

Prev = 11110000
Curr = 00001111
XOR = 11111111 ⇒ HD = 8 (many toggles)
Weight = 1111 0000 ⇒ popcount(upper) = 4 (many upper bits)
Combined ⇒ High HD + many upper bits ⇒ longer path (slow)

Prev = 11110000
Curr = 11110001
XOR = 00000001 ⇒ HD = 1 (few toggles)
Weight = 0000 1111 ⇒ popcount(upper) = 0 (few upper bits)
Combined ⇒ Low HD + few upper bits ⇒ shorter path (fast)

**Fig. 3.** *Tier-classification examples: high activation toggling triggers HD–Slow, while dense upper-half bits trigger MSB–Slow; Hybrid asserts Slow only when both features are high, otherwise selecting Mid/Fast.*

or re-execution. The overall systolic rhythm remains aligned, ensuring seamless data propagation across the array while limiting throughput loss to infrequent Mid and Slow hold events.

To capture the multimodal slack behavior observed in Figure 1, we use three timing tiers. Two tiers are too coarse to follow distinct slack modes, while four or more introduce unnecessary control complexity. The three-tier configuration is chosen to capture the dominant modes observed in the measured slack distribution (Figure 1). Most operations finish well before the clock edge, while a small subset operate close to or slightly beyond it. These natural groupings correspond to fast, typical, and slow timing regimes, providing an effective trade-off between accuracy, simplicity, and hardware cost.

Figure 2 provides a high-level view of the proposed DSAC flow, illustrating the HD and MSB predictor architectures. The Hybrid predictor, which fuses both features for unified tier selection and control, is detailed in Figure 4.

## 3.2 Scope and Dataflow Assumption

We assume a weight-stationary dataflow, where weights remain fixed within each processing element (PE) while activations stream horizontally across cycles. This mapping minimizes memory bandwidth and data movement energy, consistent with the design of modern TPU-style systolic arrays. Under this mode, delay variation is dominated by temporal changes in the activation stream, since weights are constant and only the activation inputs toggle across cycles.

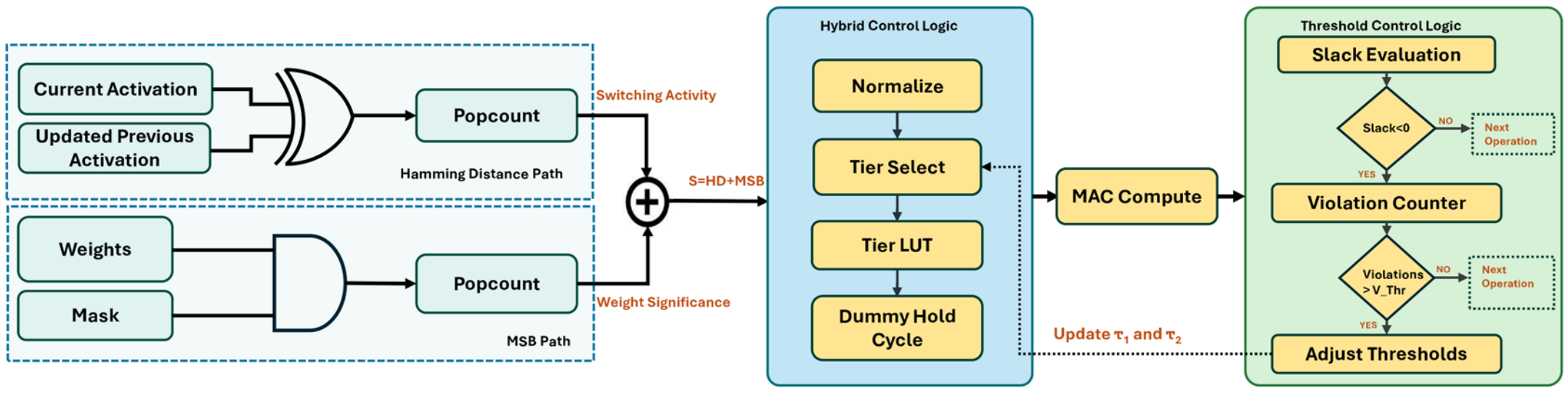

**Fig. 4.** *Hardware implementation of the Hybrid predictor. Combines HD and MSB inputs to map MACs into three timing tiers. This integration supports up to 1.37× better energy efficiency by selectively extending only the most critical execution windows.*

Consequently, the temporal switching term in the HD and Hybrid predictors is computed between consecutive activations (`Prev_act` → `Curr_act`).

The weight-stationary configuration dictates that delay and timing slack variation primarily occur along the activation dataflow dimension. Successive activation toggles produce correlated switching activity across columns of the array, leading to predictable temporal variation in MAC delays. This correlation enables the potential reuse of tier decisions across neighboring PEs or cycles, further reducing per-cycle control overhead without compromising accuracy. The MSB and Hybrid predictors, however, are dataflow-agnostic: they depend only on operand bit patterns rather than the direction of data movement. In a weight-stationary array, the MSB term naturally associates with the weight operand, but in other mappings—such as output-stationary or activation-stationary dataflows—the same formulation applies by assigning the magnitude term to the stationary operand and the switching term to the operand that varies across cycles.

Our implementation and analysis focus on the weight-stationary configuration for clarity and comparability with prior TPU-based designs; however, the proposed predictor framework remains general. Extending it to other dataflows only requires identifying which operand exhibits dominant temporal transitions and assigning the HD component accordingly. This flexibility ensures that the overall DSAC mechanism remains applicable across diverse accelerator architectures, independent of whether timing budgets are enforced using dummy-hold cycles on a fixed fast clock or another equivalent local stretching mechanism.

### 3.3 Why These Predictors?

Near-threshold timing variation arises from two dominant data-dependent effects: (1)temporal switching activity, which governs instantaneous power draw and momentary supply fluctuations, and (2)operand magnitude, which determines how deeply a computation propagates through the multiplier's partial-product and carry-propagation trees. Each proposed predictor targets one of these physical causes—or both in the Hybrid case—to expose fine-grained timing diversity without requiring any architectural modifications. DSAC uses these predictor scores to select a Fast/Mid/Slow tier, which determines the effective execution budget via the tier LUT and dummy-hold control.

**HD predictor:** When an activation word changes drastically between consecutive cycles, a large fraction of internal nodes switch simultaneously. This surge in transient current momentarily lowers the local supply voltage and increases the gate delay of nearby paths. Conversely, if few bits toggle, most internal capacitances stay quiet and signal transitions complete faster. The HD metric therefore estimates instantaneous dynamic load by measuring the fraction of bits that flip between two cycles. High toggle density implies a heavier capacitive load and longer delay,

**MSB predictor:** Even without many bits switching, some data naturally uses longer logic paths. In a multiplier, larger numbers activate more partial-product rows and longer carry chains. At low voltages, the delay increases very quickly as these logic paths get deeper. By counting the set bits in the upper half of each number, the MSB predictor identifies these structural delays. A high MSB count shows that the unit is processing a large value and will likely be slower, while a low count means the path is short and fast.

**Hybrid predictor:** In many real workloads, both sources of delay variation happen at the same time: a large number may also cause many bits to flip compared to the previous cycle. These cycles experience high electrical load and long logic paths simultaneously, which increases the total delay. The Hybrid predictor combines these two factors into a single score that matches the real delay. Using $\alpha_t$ for the bit-switching (HD) score and $\mathrm{UH}_t$ for the operand size (MSB) score, the formula is:

$$H_t = \frac{\alpha_t + 0.5\,\mathrm{UH}_t}{1.5}$$

This weighting gives more importance to the frequent bit-switching ($\alpha_t$) while still considering the size of the data ($\mathrm{UH}_t$). This linear combination closely follows the real delay trends found in our timing analysis and is very efficient to build in hardware.

To see how this works, consider two 8-bit examples (Fig. 3): (1) When many bits flip and most upper bits are "1", both predictors show high values. This results in a high Hybrid score and a **Slow-tier** assignment. (2) When only one bit flips and the upper bits are zero, the score is very low, which selects the **Fast tier**. Other data patterns naturally fall into the **Mid tier**. This behavior shows how delay changes with the data: frequent switching increases the electrical load, while larger numbers activate longer logic paths. By using these predictors together, the system can adjust its timing for every cycle based on the actual data being processed.

### 3.4 Predictor Design

In this section, we present the mathematical formulation and circuit implementation of the HD, MSB, and Hybrid predictors. Each predictor is computationally lightweight and produces a normalized criticality score $s_t \in [0,1]$ used for per-operation tier assignment. We then describe how this score is mapped to timing tiers and enforced via local dummy-hold control.

***(a) Hamming-Distance (HD) Predictor.*** The HD predictor computes a per-cycle switching score from the activation stream and converts it into a normalized value used for tier selection. It measures the bit-level difference between consecutive activations and produces the toggle ratio $\alpha_t \in [0,1]$.

Each processing element (PE) keeps a one-cycle-delayed copy of its last activation, denoted as `Prev_act`. When the next activation (`Curr_act`) arrives, the PE performs a bitwise comparison between the two. The resulting metric, called the normalized toggle ratio, quantifies how many bits flipped and is defined as

$$\alpha_t = \frac{1}{d}\sum_{i=0}^{d-1} \mathbf{1}\{\mathrm{Prev_act}[i] \neq \mathrm{Curr_act}[i]\},$$

where $d$ is the operand width (8 in this work) and $1\{\cdot\}$ is the indicator function (1 if true, 0 otherwise). The numerator counts the number of bit positions that changed between cycles, and dividing by $d$ scales the count to a normalized score in [0,1]. As illustrated in Fig. 3, higher toggle counts between `Prev_act` and `Curr_act` produce higher $\alpha_t$ and map to the Slow tier, while lower toggle counts map to the Fast tier, with intermediate cases mapping to the Mid tier.

The hardware implementation of the HD predictor integrates seamlessly into the proposed DSAC framework, as visually summarized in the left panel of Figure 2. Within each processing element, the activation stream enters alongside the locally stored `Prev_act` value. An XOR array computes bitwise differences to generate a toggle mask, which then feeds a compact popcount tree that aggregates the total number of bit flips. This result represents the number of activation-bit toggles in the current cycle. The normalized toggle ratio, $\alpha_t$, is then compared against two programmable thresholds in the Tier Select unit to classify the operation as Fast, Mid, or Slow. The selected tier index is passed to the Tier-LUT, which returns the corresponding dummy-hold count that defines the cycle's effective execution window. This localized, data-dependent control allows each MAC to adapt its timing budget dynamically without changing the fast clock or interrupting the overall systolic flow. After applying the dummy-hold cycles, the PE executes the MAC operation using the extended execution window and then resumes normal data propagation. The Threshold Control Logic monitors timing outcomes (violations) and updates the tier thresholds ($\tau_1$,$\tau_2$) when the violation count exceeds a programmable limit $V_{\text{thr}}$, enabling closed-loop retuning under different workloads.

Expensive division hardware is avoided by absorbing normalization directly into the comparator thresholds. Both the XOR and popcount structures are shallow enough to meet timing requirements comfortably at the base frequency, while the comparator and tier encoder add only negligible delay to the critical path. The `Prev_act` register updates each cycle under normal operation, and stalls or pipeline bubbles are masked to prevent spurious toggles. By translating raw per-cycle bit activity into a quantitative score that tracks real delay behavior, the HD predictor provides a fine-grained, low-cost mechanism for identifying cycles likely to require an extended execution window. This enables the TPU array to selectively extend only critical cycles while allowing non-critical ones to proceed without added delay, improving both timing resilience and energy efficiency.

***(b) MSB Predictor.*** The MSB predictor estimates per-operation timing criticality from operand magnitude. In the weight-stationary dataflow evaluated in this work, the predictor is applied to the stationary weight operand; however, the same logic can be applied to the operand whose magnitude most strongly influences the multiplier delay in other dataflows. The key observation is that operands with larger high-order magnitude components tend to activate deeper partial-product and carry-propagation paths in the multiplier, especially under near-threshold operation where delay is highly sensitive to logic depth.

For the INT8 configuration used in our main evaluation, the predictor inspects the upper half of the operand and computes a normalized upper-bit population score. Let $x$ denote the selected operand and let $d$ be its bit-width. The MSB score is defined as

$$\text{MSB_score} = \frac{\text{popcount}(x[d-1{:}\,d/2])}{d/2},$$

where $x[d-1{:}d/2]$ selects the most-significant half of the operand and `popcount` counts the number of set bits in that field. For $d$=8, this corresponds to bits [7:4]. Thus, an operand such as `11110000` has four set bits in the upper half and produces MSB_score=1.0, which typically maps to the Slow tier. In contrast, `00110000` produces MSB_score=0.5 and is mapped according to the configured tier thresholds. Higher MSB scores therefore indicate larger operand magnitude and a higher likelihood of requiring an extended execution window.

**Precision-scalable formulation.** Although the main evaluation uses INT8 operands, the MSB predictor is not tied to a fixed 8-bit representation. For an integer operand with bit-width $b$, the predictor can be generalized by selecting a configurable number of upper magnitude bits:

$$\text{MSB_score}_{\text{int}} = \frac{\text{popcount}(x[b-1{:}\,b-k])}{k},$$

where $k$ is the number of upper magnitude bits used for prediction. In this work, $k$=$b$/2 is used for consistency across integer precisions. This reduces to the upper four bits for INT8 and the upper two bits for INT4. For signed integer formats, the sign bit is excluded and the score is computed over the magnitude field so that the predictor captures operand size rather than sign polarity. Only the mask width and threshold values need to be changed; the hardware structure remains the same, consisting of a bit mask, a small popcount tree, and two programmable comparators.

For floating-point formats such as FP8, the upper bits do not represent integer magnitude directly. Therefore, the MSB predictor is reformulated as a format-aware magnitude predictor. In FP8, the exponent field determines the numerical scale of the operand and is the closest analogue to the integer MSB region. The magnitude score can therefore be computed from the normalized exponent value, optionally combined with selected leading mantissa bits:

$$\text{MSB_score}_{\text{fp}} = \alpha \cdot \frac{E - E_{\text{min}}}{E_{\text{max}} - E_{\text{min}}} + (1 - \alpha) \cdot \frac{\text{popcount}(M_{\text{lead}})}{|M_{\text{lead}}|},$$

where $E$ is the exponent value, $E_{\min}$ and $E_{\max}$ define the valid exponent range, $M_{\text{lead}}$ denotes the selected leading mantissa bits, and $\alpha$ controls the relative contribution of exponent and mantissa information. This distinction is important because, unlike INT8, the most significant FP8 bits encode scale rather than a direct binary integer magnitude. The exponent term captures coarse operand scale, while the optional mantissa term captures variation in the significand multiplication path. The parameter $\alpha$ and the tier thresholds can be calibrated using the same STA-based delay-correlation flow used for INT8.

This format-aware definition preserves the role of the MSB predictor as a lightweight proxy for operand magnitude and timing criticality, while allowing the same Tier-LUT and threshold-control mechanism to support INT8, INT4, and FP8 operands. In practice, only the field-selection logic and tier thresholds change across formats: integer formats use upper magnitude bits, whereas FP8 uses the exponent field and, optionally, selected leading mantissa bits.

Figure 2 (right panel) illustrates the MSB predictor datapath and its integration with the timing-control pipeline for the integer configuration evaluated in this work. The incoming operand is first masked to isolate the upper-half bits, followed by a compact popcount tree that computes the number of set bits. The resulting count is compared against two programmable thresholds in the Tier Select unit to classify the operation into Fast, Mid, or Slow. The tier index is forwarded to the Tier-LUT, which returns the corresponding dummy-hold count that defines the cycle's effective execution window. The dummy-hold mechanism enforces the LUT decision by stalling local pipeline state for the specified number of cycles before the MAC operation proceeds. Finally, the threshold-control logic monitors timing outcomes (violations) and can update the tier thresholds ($\tau_1$,$\tau_2$) when the violation count exceeds a programmable limit $V_{\text{thr}}$, enabling closed-loop retuning under different workloads.

The MSB predictor is implemented as a small, low-overhead datapath that operates off the main MAC critical path. The mask, field-extraction, popcount, and comparison structures are shallow and meet timing comfortably at the base frequency. Unlike the HD predictor, the MSB datapath is purely combinational and does not require a previous-cycle register, since the score depends only on the current operand's selected magnitude field: upper magnitude bits for integer formats, or exponent and optional leading mantissa bits for FP8 formats. By translating this selected magnitude field into a normalized score that correlates with operand-dependent logic depth, the MSB predictor provides a low-cost mechanism for identifying cycles likely to require an extended execution window. This enables the TPU array to selectively extend only magnitude-critical cycles while allowing non-critical ones to proceed without added delay, improving both timing resilience and energy efficiency.

***(c) Hybrid Predictor.*** The Hybrid predictor combines the HD-based switching score and the MSB-based magnitude score into a single normalized criticality metric used for tier selection. For each MAC operation at cycle $t$, it computes a switching activity term and a format-aware magnitude term:

$$\mathrm{HD}_t = \frac{\text{XOR toggles}}{b_a}, \qquad \mathrm{MAG}_t = \mathcal{M}(w_t, \mathcal{F}),$$

where $b_a$ is the activation bit-width, $w_t$ is the selected weight operand, and F denotes the numeric format. Here, $\mathrm{HD}_t$ captures the normalized Hamming-distance activity between consecutive activations, while $\mathrm{MAG}_t$ captures the format-aware operand-magnitude information produced by the MSB predictor of Section 3.4(b). For integer formats, $b_a$ is the operand bit-width and HD is computed over all activation bits. For FP8 formats, HD is computed over the full 8-bit representation; refining HD to weight specific FP8 fields, such as mantissa toggles only, is left as future work.

For the INT8 configuration evaluated in this work, $b_a$=8 and the magnitude term reduces to the original upper-half score:

$$\mathrm{MAG}_t = \mathrm{UH}_t = \frac{\text{popcount(weight[7:4])}}{4}.$$

For INT4, the same logic uses the upper two magnitude bits with retuned thresholds. For FP8, the integer upper-bit popcount is replaced by an exponent-based magnitude term, optionally combined with selected leading mantissa bits, because the exponent captures the numerical scale of the FP8 operand. Thus, the Hybrid predictor preserves the same control structure while allowing the magnitude feature to be adapted to the operand precision and numeric format.

These two features are fused into a single score:

$$H_t = \frac{\mathrm{HD}_t + 0.5 \times \mathrm{MAG}_t}{1.5},$$

where the factor 0.5 down-weights the magnitude term and the denominator 1.5 re-normalizes the combined metric to the [0,1] range. In the INT8 case, $\mathrm{MAG}_t$=$\mathrm{UH}_t$, so the expression reduces to the original Hybrid formulation:

$$H_t = \frac{\mathrm{HD}_t + 0.5 \times \mathrm{UH}_t}{1.5}.$$

A higher $H_t$ indicates a more critical operation and therefore a need for a larger effective timing budget (execution window), while a lower $H_t$ corresponds to a less critical operation.

Figure 4 illustrates the Hybrid datapath and its integration with the timing-control pipeline. The switching feature path computes $\mathrm{HD}_t$ using XOR and popcount between the current activation and the updated previous-activation value. In parallel, the magnitude feature path computes $\mathrm{MAG}_t$ using the format-aware MSB logic described in Section 3.4(b). For INT8, this reduces to mask-and-popcount on the weight operand's upper-half bits; for FP8, the exponent field is extracted instead. The two features are combined to form $H_t$ and compared against two programmable thresholds to classify the cycle into Fast, Mid, or Slow. The selected tier index is forwarded to the Tier-LUT, which returns the corresponding dummy-hold count that defines the cycle's effective execution window. The dummy-hold mechanism then stalls local pipeline state for the specified number of cycles before the MAC operation proceeds.

The same figure also shows the feedback path used for closed-loop retuning. After MAC completion, the control logic evaluates slack using $T_{\mathrm{eff}}$; if slack is negative, the design registers a timing violation and increments a violation counter. When the counter exceeds a programmable limit $V_{\mathrm{thr}}$, the controller updates the tier thresholds ($\tau_1$,$\tau_2$) and continues operation, enabling adaptation under different workload changes.

The Hybrid predictor reuses the same XOR/popcount and format-aware magnitude structures as the standalone HD and MSB predictors respectively, adding only a small fusion step and tier-selection logic on the control path. By combining switching and magnitude-driven features into a single score, the Hybrid predictor provides a robust mechanism for identifying the most timing-critical cycles and selectively extending only those cycles via dummy-hold insertion, while allowing non-critical cycles to proceed without added delay, improving both timing resilience and energy efficiency.

### 3.5 Tier Mapping and Thresholding

Let $s_t \in \{\alpha_t, \text{MSB_score}, H_t\}$ denote the predictor output and $(\tau_1, \tau_2)$ be programmable thresholds:

$$s_t < \tau_1 \Rightarrow \text{Fast}, \quad \tau_1 \leq s_t < \tau_2 \Rightarrow \text{Mid}, \quad s_t \geq \tau_2 \Rightarrow \text{Slow}.$$

After classification, DSAC assigns a dummy-hold count $h$ to enforce the selected tier under a fixed fast clock with period $T_{\text{fast}}$. In our three-tier configuration, Fast, Mid, and Slow correspond to $h \in \{0,1,2\}$ dummy holds, respectively, yielding an effective timing budget

$$T_{\text{eff}} = (1 + h)\, T_{\text{fast}}.$$

Thus, Fast operations execute in one fast-clock cycle, while Mid and Slow operations are given additional time by stalling local pipeline state for one or two extra fast-clock cycles before allowing the systolic wave to advance.

The thresholds $(\tau_1, \tau_2)$ are configurable parameters that define the Fast/Mid/Slow boundaries and can be tuned offline to match the observed delay distribution, or adjusted at runtime by the feedback controller when the accumulated violation count exceeds a programmable limit $V_{\text{thr}}$. This allows DSAC to balance performance and timing resilience under workload changes without manual retuning. Predictor normalization is handled implicitly through the threshold values, avoiding explicit normalization hardware while preserving functional equivalence.

| Benchmarks | | Error free Accuracy |
|---|---|---|
| **Name** | **Basic Layers Architecture** | |
| SVHN [13] | CONV: (32, 32, 3)x(32, 32, 32)x(32, 32, 32)x(14, 14, 64)x(14, 14, 64)x(5, 5, 128)x(5, 5, 128), FC: 512x512x10 | 0.94 |
| GTSRB [14] | CONV: (3, 48, 48)x(32, 48, 48)x(32, 46, 46)x(64, 23, 23)x(64, 21, 21)x(128, 10, 10)x(128, 8, 8), FC: 2048x512x43 | 0.97 |
| Reuters [15] | FC: 2048x256x256x46 | 0.80 |
| IMDB [16] | CONV: 400x(400x50)x(398, 256), FC: 256x1 | 0.89 |
| MNIST [17] | FC: 784x256x256x10 | 0.98 |
| CIFAR-10 [18] | CONV: (32, 32, 3)x(32, 32, 32)x(32, 32, 32)x(16, 16, 64)x(16, 16, 64)x(8, 8, 128)x(8, 8, 128), FC: 2048x512x10 | 0.77 |
| FMNIST [19] | FC: 784x256x512x10 | 0.89 |
| AMNIST [20] | CONV: (20, 25, 1)x(20,25,128)x(20,25,64), FC: 32000x256x128x40 | 0.92 |
| BERT-tiny[21] | FC: (128x128)x4 + 128x512 + 512x128 (x2 layers), 128x128, 128x2 | 0.86 |
| ResNet-50[22] | CONV: ResNet-50 backbone (50 layers), FC: 2048x1000 | 0.66 |

**Table 1.** *List of ten benchmarks used and their error free accuracy.*

### 3.6 Tier Selection

Each MAC stores the 2-bit tier index produced by the tier-selection logic, which is translated into the dummy-hold count $h$ defined in Section 3.5. The PE then stalls its local pipeline state for $h$ additional $T_{\text{fast}}$ cycles before allowing the systolic wave to advance, keeping the global fast clock unchanged. Figure 2 summarizes this end-to-end control flow: predictor feature extraction and tier selection produce a tier index, which drives the dummy-hold block before `MAC Compute`. The thresholds $(\tau_1, \tau_2)$ remain configurable, allowing designers to tune tier boundaries either in simulation or post-silicon to meet a desired timing-violation target. This flexibility enables DSAC to sustain high energy efficiency and error resilience across varying process, voltage, and temperature (PVT) conditions.

**Algorithm 1** DSAC Control Flow

```
1:  Input: Activation A_t, Weight W_t, tier thresholds (τ1, τ2), fast clock period T_fast
2:  Initialize violation counter: viol_cnt ← 0
3:  Compute predictor score s_t using HD, MSB, or Hybrid logic
4:  if s_t < τ1 then
5:      Assign Fast tier; set h ← 0
6:  else if τ1 ≤ s_t < τ2 then
7:      Assign Mid tier; set h ← 1
8:  else
9:      Assign Slow tier; set h ← 2
10: end if
11: Apply dummy-hold control for h fast-clock cycles
12: Execute MAC operation with effective budget T_eff = (1 + h) T_fast
13: if violation then
14:     viol_cnt ← viol_cnt + 1
15: end if
16: if viol_cnt > V_thr then
17:     Update (τ1, τ2)
18:     viol_cnt ← 0
19: end if
20: Continue to next MAC operation
```

To clearly describe the control sequence executed within each processing element, Algorithm 1 outlines the per-operation control flow of DSAC. The procedure shows how each MAC computes its predictor score, maps the result to a timing tier, and applies dummy-hold cycles according to the selected tier.

Algorithm 1 represents the control sequence executed within each processing element. Lines 1–2 define the inputs and configuration state: the current activation and weight operands, the tier thresholds ($\tau_1$,$\tau_2$) that partition the predictor score into Fast, Mid, and Slow regions, and the fixed fast-clock period $T_{\text{fast}}$. Line 3 computes the predictor score $s_t$ using the selected predictor (HD, MSB, or Hybrid). Lines 4–10 map $s_t$ to a timing tier via the two-threshold rule and assign the corresponding dummy-hold count $h\in\{0,1,2\}$. Lines 11–12 enforce the selected timing budget by stalling the PE's local pipeline state for $h$ fast-clock cycles, yielding $T_{\text{eff}}$=(1+$h$) $T_{\text{fast}}$, before completing the MAC operation and propagating its result. Lines 13–19 implement the feedback mechanism: when a timing violation is detected, the violation counter is incremented, and if the counter exceeds $V_{\text{thr}}$ the thresholds ($\tau_1$,$\tau_2$) are updated and the counter is reset. This closed-loop adjustment allows the tier boundaries to be retuned during execution to maintain timing safety under workload and PVT variation while preserving the overall systolic operation.

### 3.7 Tier Lookup Table (LUT)

The Tier Lookup Table (Tier-LUT) serves as the interface between the predictor output and the timing-control hardware, converting qualitative tier decisions into quantitative timing adjustments. Once a tier index is generated, each MAC queries this small LUT to obtain a dummy-hold count for the selected tier. Under DSAC, the array operates from a single fixed fast clock with period $T_{\text{fast}}$, and the effective timing budget is realized by stalling for additional fast-clock cycles:

$$T_{\text{eff}} = (1 + h)\, T_{\text{fast}}.$$

where $h_i$ denotes the number of local dummy-hold cycles inserted for tier $i$. These holds are implemented using ready/valid gating (stalling local pipeline state) while keeping the fast clock unchanged, ensuring that the array remains fully synchronous. A larger $h_i$ value thus provides extra settling time only for timing-critical MAC operations, improving timing safety without requiring global DVFS or replay.

Entries in the Tier-LUT are derived from the delay distribution obtained through STA and can be updated to reflect post-layout characterization or calibration. The LUT requires only three entries (Fast, Mid, Slow) and can be shared across a small PE cluster, keeping control overhead minimal.

### 3.8 Dummy-Hold Cycle Implementation

The dummy-hold mechanism realizes the timing extensions specified by the Tier-LUT. When a PE receives a Mid- or Slow-tier decision, it asserts hold_req to stall its local pipeline registers for $h$ additional fast-clock cycles, allowing the current MAC operation extra time to complete. Therefore, DSAC does not assume that every MAC operation completes within the Fast-tier timing budget. Operations that are predicted to require a longer settling time are assigned to the Mid or Slow tier, where the effective timing budget is extended by one or two additional fast-clock ticks, respectively.

Figure 5 illustrates this behavior at the systolic dataflow level. In the Fast tier, no hold is inserted, and the active wavefront advances normally: activations propagate horizontally across columns, while partial sums propagate vertically across rows. In contrast, a Mid-tier decision holds the active wavefront for one dummy tick, and a Slow-tier decision holds it for two dummy ticks. During these hold ticks, the activation and partial-sum state remains stable and downstream propagation is delayed in a controlled manner.

For MAC operations that are not early-finishing, DSAC intentionally trades additional local execution time for timing safety. Fast-tier operations incur no extra cycles, while Mid- and Slow-tier operations introduce one and two dummy-hold cycles, respectively. These extra cycles increase the physical execution ticks for the corresponding timing-critical operations, but they avoid timing violations by keeping the activation and partial-sum state stable until the MAC result can be safely propagated. The resulting performance overhead is proportional to the fraction of operations classified as Mid or Slow, and is captured at the architecture level through the FAST/MID/SLOW tier counts, dummy-hold counts, physical ticks, and normalized throughput.

To maintain spatial alignment of the systolic wavefront, DSAC integrates the dummy-hold mechanism with local stall control so that neighboring PEs observe temporally consistent operands. In the weight-stationary dataflow, activations advance horizontally while partial sums propagate vertically; therefore, hold cycles must prevent premature forwarding of either value while a timing-critical MAC is being extended. Overall, dummy-hold insertion trades a bounded amount of local hold-cycle overhead for improved timing safety under aggressive near-threshold operation, without changing the global fast reference clock or requiring DVFS or replay-based recovery.

### 3.9 Threshold Control Logic

While the tier thresholds ($\tau_1$,$\tau_2$) can be statically tuned offline, DSAC also supports a lightweight feedback loop to maintain timing safety under workload and PVT variation. After each MAC completes, the control logic evaluates the observed slack,

$$\text{slack}_t = T_{\text{eff}} - D_t,$$

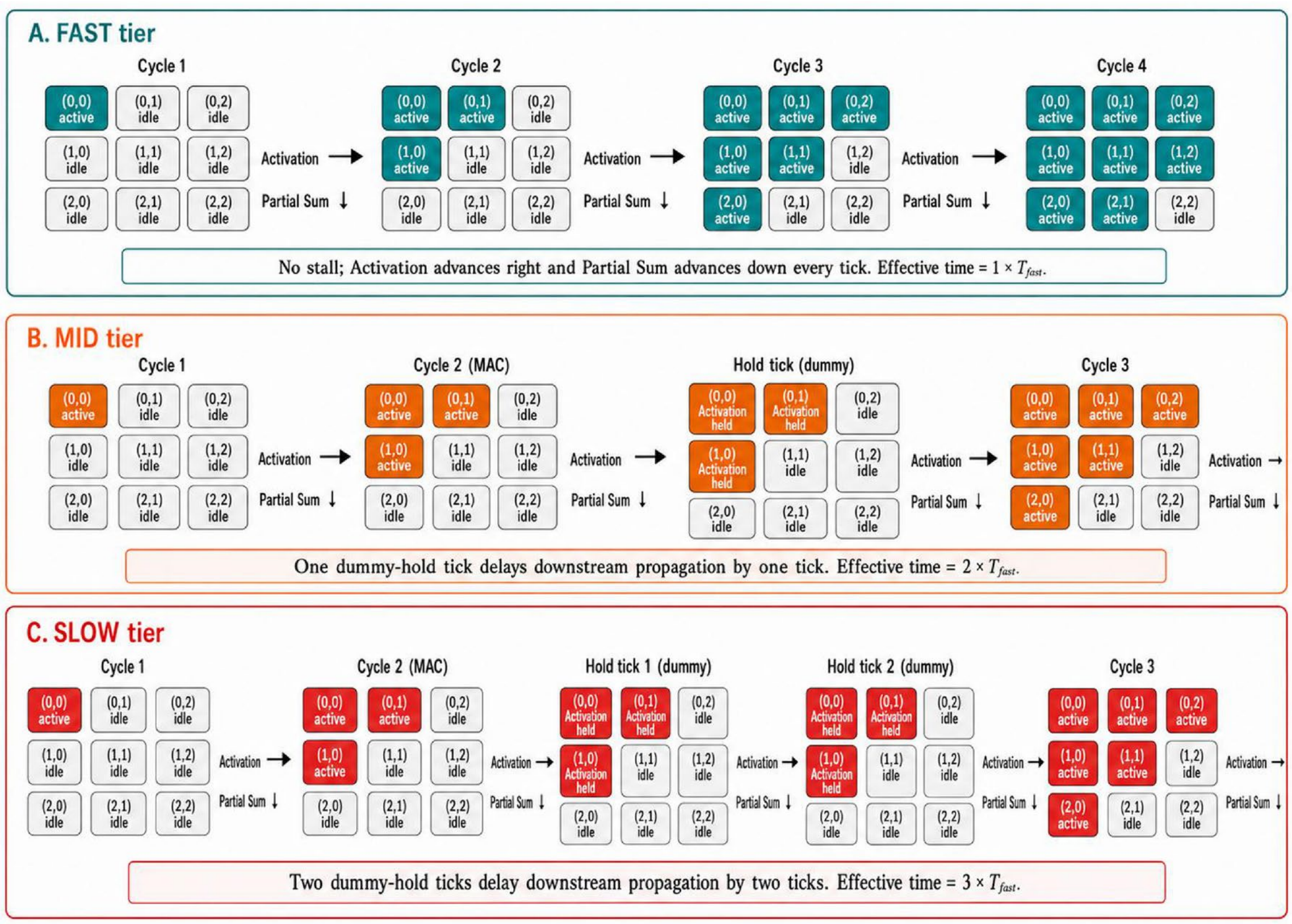

**Fig. 5.** *Cycle-level effect of DSAC dummy-hold control on a weight-stationary systolic TPU dataflow. Activations propagate horizontally while partial sums propagate vertically. Fast-tier operations advance every fast-clock tick, whereas Mid- and Slow-tier operations hold the active wavefront state for one and two dummy ticks, respectively. These controlled holds delay downstream propagation and increase physical execution ticks for timing-critical operations, while preserving timing correctness under a fixed global fast reference clock.*

where $D_t$ is the MAC delay and $T_{\text{eff}}=(1+h)\,T_{\text{fast}}$ is the effective timing budget realized by the selected tier and dummy-hold count. A negative slack indicates that the operation would violate the selected timing budget; DSAC flags this event as a timing violation.

To prevent over-correction or permanent performance degradation, the threshold controller monitors the violation rate $r$ over a programmable window and applies the bi-directional logic detailed in Algorithm 2. When the violation rate falls below a lower bound $y$, the controller performs an aggressive recovery by incrementing the thresholds ($\tau_1,\tau_2$) by $\Delta_2$. This effectively loosens the classification, allowing more operations to be processed in the Fast tier to maximize throughput. Conversely, if the violation rate exceeds the safety limit $x$, the controller initiates a conservative back-off by shifting the thresholds downward by $\Delta_1$. This forces a larger percentage of operations into the Mid and Slow tiers, thereby increasing the timing margin and ensuring safety. This dual-action tuning provides robustness to operating-point drift. As shown in Fig. 9, when the normalized frequency increases, the thresholds automatically tighten to reclassify cycles into safer tiers.

**Algorithm 2** Two-Check Threshold Update

**Require:** Violation rate $r$, bounds $x > y$, step sizes $\Delta_1$, $\Delta_2$
**Require:** Current thresholds $\tau_1, \tau_2$
1: **if** $r > x$ **then**
2: $\tau_1 \leftarrow \tau_1 - \Delta_1$
3: $\tau_2 \leftarrow \tau_2 - \Delta_1$
4: **else if** $r < y$ **then**
5: $\tau_1 \leftarrow \tau_1 + \Delta_2$
6: $\tau_2 \leftarrow \tau_2 + \Delta_2$
7: **end if**

Once operating conditions stabilize or the frequency drops, the two-check mechanism allows the thresholds to recover, maintaining a balance between performance and timing safety.

DSAC utilizes a hysteresis-based approach to prevent rapid threshold oscillations and ensure system stability. By requiring $V_{\text{thr}}$ violations before an update, the controller effectively filters out transient noise or isolated data-dependent delay spikes. Furthermore, the use of small, incremental step sizes ($\Delta_1$,$\Delta_2$) ensures that the system converges toward an optimal threshold boundary gradually rather than over-correcting. These safeguards, combined with optional update-interval enforcement, ensure the framework remains robust to operating-point drift without triggering the ping-pong effect between timing tiers. Overall, the feedback and threshold-control logic adds minimal hardware overhead—primarily a small counter and subtract logic—while improving resilience by adapting tier selection to the observed timing behavior without changing the fast clock. A more exhaustive sensitivity study of ($\Delta_1$,$\Delta_2$,$V_{\text{thr}}$) is left for future work.

## 4 Experimental Methodology

This section presents the cross-layer methodology used to design and evaluate DSAC. Our evaluation flow spans four abstraction levels—device, circuit, architecture, and application—capturing how near-threshold variability propagates from low-level gate-delay variation to system-level throughput, energy efficiency, and end-to-end inference accuracy [6, 5]. Figure 7 later reports the resulting application-level accuracy trends, while Figure 6 summarizes the complete validation flow across the four layers.

**Cross-layer validation flow.** Figure 6 summarizes the cross-layer validation methodology used to evaluate DSAC across the device, circuit, architecture, and application layers. The validation begins at the device layer, where near-threshold, variation-aware gate-delay distributions are generated using PTM, VARIUS-NTV, and VARIUS-TC models. These device-level delay models are used to construct gate-delay libraries for circuit-level analysis.

At the circuit layer, the synthesized MAC RTL/netlist and operand vectors are analyzed using Synopsys DC and our in-house STA engine to produce operand-dependent per-MAC delay values $D_t$, slack distributions, and timing-violation information. The resulting delay traces are then passed to the architecture layer, where the cycle-accurate TPU systolic-array simulator evaluates DSAC tier selection, dummy-hold insertion, FAST/MID/SLOW tier counts, timing violations, physical ticks, and normalized throughput. Finally, at the application layer, quantized DNN workloads extracted from Keras models are executed layer-by-layer through the simulator to measure end-to-end inference accuracy, normalized TOPS, normalized power, and TOPS/Watt. This flow connects low-level NTC delay variability directly to circuit-level MAC timing, architecture-level DSAC behavior, and application-level correctness and efficiency.

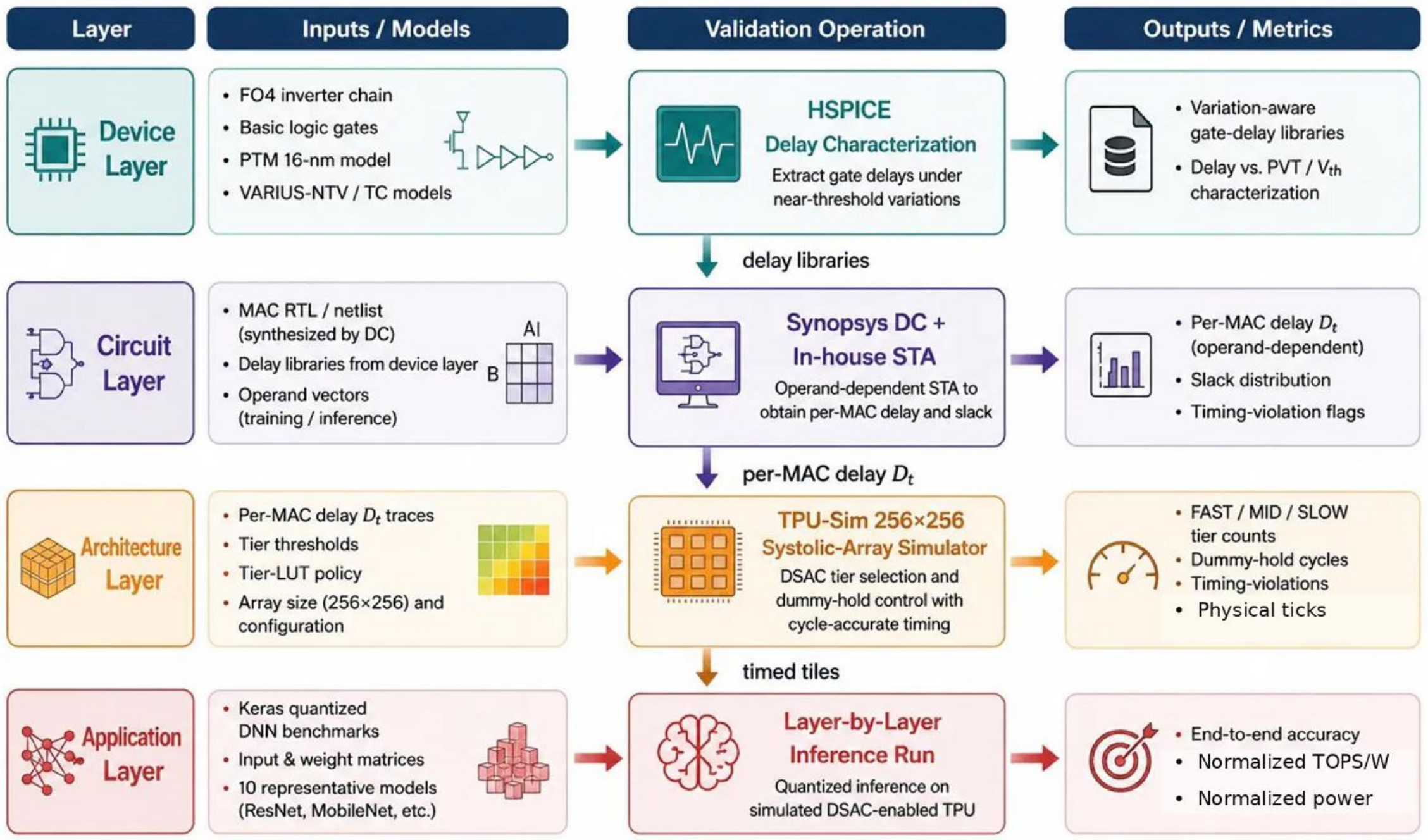

**Fig. 6.** *Cross-layer experimental validation methodology for DSAC, spanning device, circuit, architecture, and application layers. Each layer feeds its output to the next: device-level delay libraries inform circuit-level STA, which produces per-MAC delays consumed by the architecture-level systolic simulator, ultimately yielding application-level accuracy and efficiency metrics.*

**Simulation tools and analysis settings.** The experimental setup combines device-level delay modeling, circuit-level timing analysis, architecture-level simulation, and application-level inference evaluation. At the device level, near-threshold gate-delay behavior is modeled using the 16 nm PTM model together with VARIUS-NTV and VARIUS-TC, which capture voltage sensitivity and process-induced delay variation. These delay models are used to support the circuit-level timing analysis of the TPU datapath.

The TPU-style MAC and systolic-array RTL are implemented in Verilog and synthesized using Synopsys Design Compiler to obtain gate-level netlists. Operand-dependent propagation delay is then evaluated using our in-house C++ STA engine, which estimates the per-MAC delay $D_t$, slack, and timing-violation behavior under process variation and input switching activity. These timing traces are integrated into an in-house cycle-accurate C++ TPU simulator, where DSAC tier selection, FAST/MID/SLOW tier counts, dummy-hold insertion, physical ticks, timing violations, normalized throughput, and power behavior are evaluated.

For application-level evaluation, trained Keras models are used to extract quantized weights, activations, and biases, which are executed layer-by-layer through the simulator. Unless otherwise stated, all results use a near-threshold baseline operating point of 0.45 V and 67.5 MHz, and frequency-scaling experiments are reported from 1.0× to 2.15× relative to this baseline. Matrix operations are partitioned into 256×256 systolic-array tiles before simulation. We evaluate Baseline, TE-Drop, HD, MSB, and Hybrid schemes under identical workload inputs and report inference accuracy, timing violations, FAST/MID/SLOW tier distributions, physical ticks, normalized TOPS, normalized power, TOPS/Watt, and area overhead.

### 4.1 Circuit Layer

At the circuit layer, we implement a TPU-style systolic array in Verilog RTL. The design is synthesized using Synopsys Design Compiler, and the resulting gate-level netlists are analyzed using our in-house STA engine developed in C++. The STA tool uses variation-aware delay libraries derived from the predictive FinFET models [23, 24, 25] to evaluate sensitized path delays under process variation and operand-dependent switching activity [6].

For each MAC operation, the STA engine reports the data-dependent propagation delay $D_t$. The delay value is compared against the baseline clock period and against DSAC's selected effective timing budget $T_{\text{eff}}$ to compute slack. Under baseline execution, slack is computed as

$$\text{slack}_{\text{base}} = T_{\text{clk}} - D_t,$$

whereas under DSAC, slack is computed as

$$\text{slack}_{\text{DSAC}} = T_{\text{eff}} - D_t.$$

These per-operation delay and slack values are consumed by the architecture simulator to determine timing violations, oracle timing tiers, and predictor classification behavior.

### 4.2 Architecture Layer

We employ an in-house TPU systolic-array simulator written in C++ and modeled on the architectural specifications of the Google TPU [26]. The cycle-accurate simulator integrates the MAC delay information generated by the STA engine (Section 4.1) to model near-threshold timing variation and data-dependent slack behavior across operations.

At this layer, each MAC operation is assigned a predictor score using one of the proposed HD, MSB, or Hybrid predictors. The score is mapped to a FAST, MID, or SLOW timing tier using programmable thresholds. The selected tier determines the number of dummy-hold cycles inserted through the Tier-LUT, thereby defining the effective timing budget

$$T_{\text{eff}} = (1 + h)T_{\text{fast}},$$

where $h \in \{0,1,2\}$ for FAST, MID, and SLOW tiers, respectively. The architecture simulator records timing violations, tier distributions, physical ticks, hold-cycle overhead, normalized throughput, and power behavior. These metrics are then used to evaluate how well DSAC converts circuit-level timing variation into architecture-level timing adaptation.

To create a realistic inference environment, the simulator is interfaced with Keras [27]. Trained DNN models are used to extract quantized activation, weight, and bias matrices, which are partitioned into 256×256 tiles and streamed through the systolic array.

### 4.3 Application Layer

At the application layer, we evaluate the system-level effect of DSAC on end-to-end DNN inference. We use the DNN benchmarks listed in Table 1. For each network, quantized activation, weight, and bias tiles are extracted from the trained model and executed layer-by-layer through the TPU simulator under identical application inputs.

The output tiles are combined to compute end-to-end inference accuracy. We compare the unprotected baseline, TE-Drop, and the proposed DSAC predictors under the same workload inputs and operating points. In addition to accuracy, we report normalized throughput, normalized power, TOPS/Watt, timing-tier distribution, and implementation overhead. This final layer closes the cross-layer validation loop by showing how device- and circuit-level delay variation affects architecture-level timing behavior and application-level quality of results.

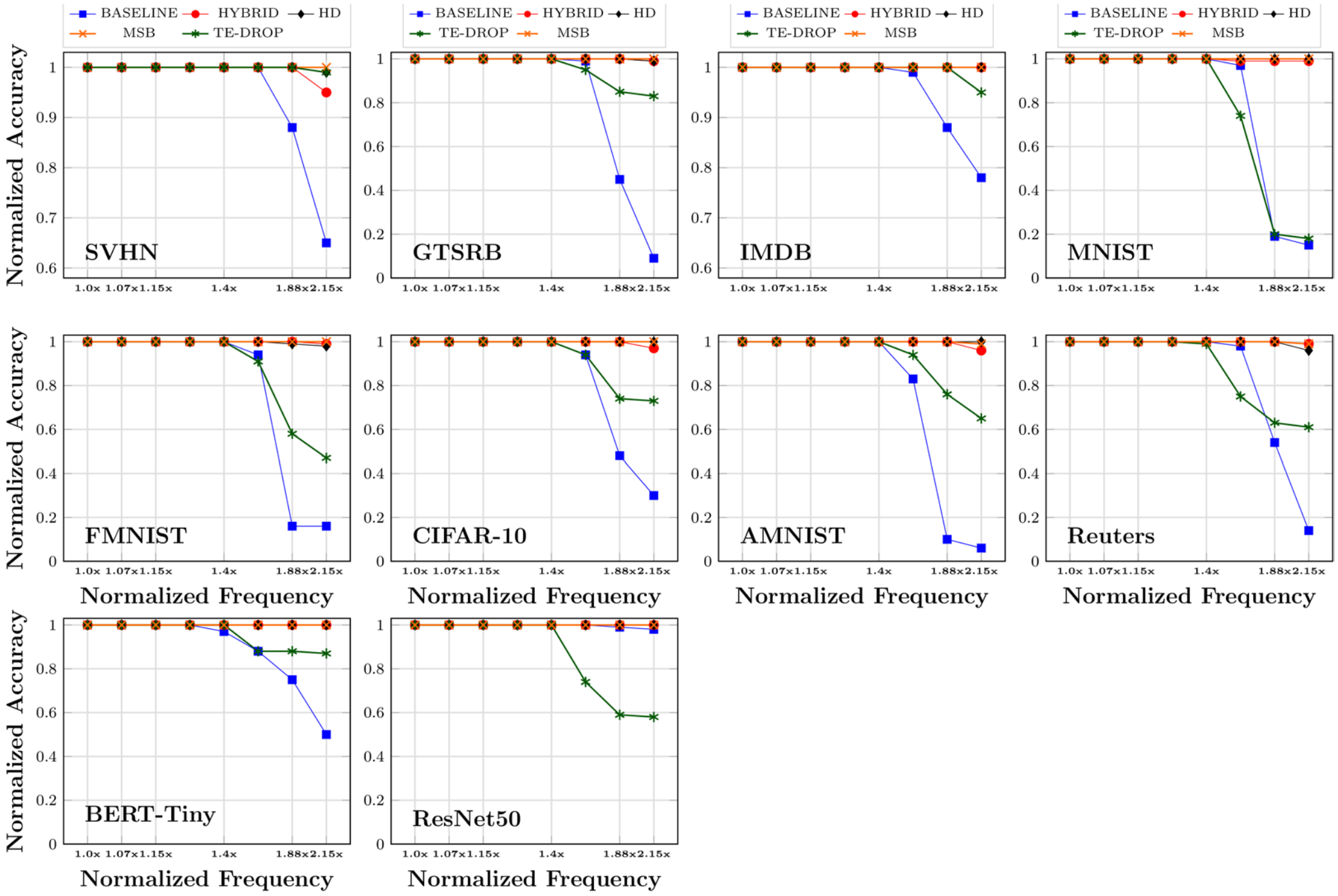

**Fig. 7.** *Normalized inference accuracies of the ten benchmarks for different comparative schemes. While the unprotected Baseline accuracy collapses, DSAC predictors maintain high stability and limit inference accuracy loss.*

## 5 Experimental Results

In this section, we evaluate the proposed DSAC framework in terms of inference accuracy, energy efficiency, and hardware overheads. Our error-free baseline operating point is 0.45 V and 67.5 MHz, selected to ensure error-free execution of the systolic array[28] across the evaluated workloads. We compare different schemes and report the resulting trends in accuracy, efficiency, and implementation overhead.

### 5.1 Comparative Schemes

We compare the following schemes in our paper:

(1) **Baseline:** This scheme executes the TPU at the baseline clock period and does not apply any timing-error mitigation.
(2) **TE-Drop:** This scheme handles timing errors by dropping the subsequent downstream MAC operation. The errant MAC steals one cycle to complete/recompute its output.
(3) **Hamming Distance (HD):** This scheme estimates timing criticality using activation bit toggles. It computes the Hamming distance between consecutive activations and maps the result to Fast/Mid/Slow tiers.

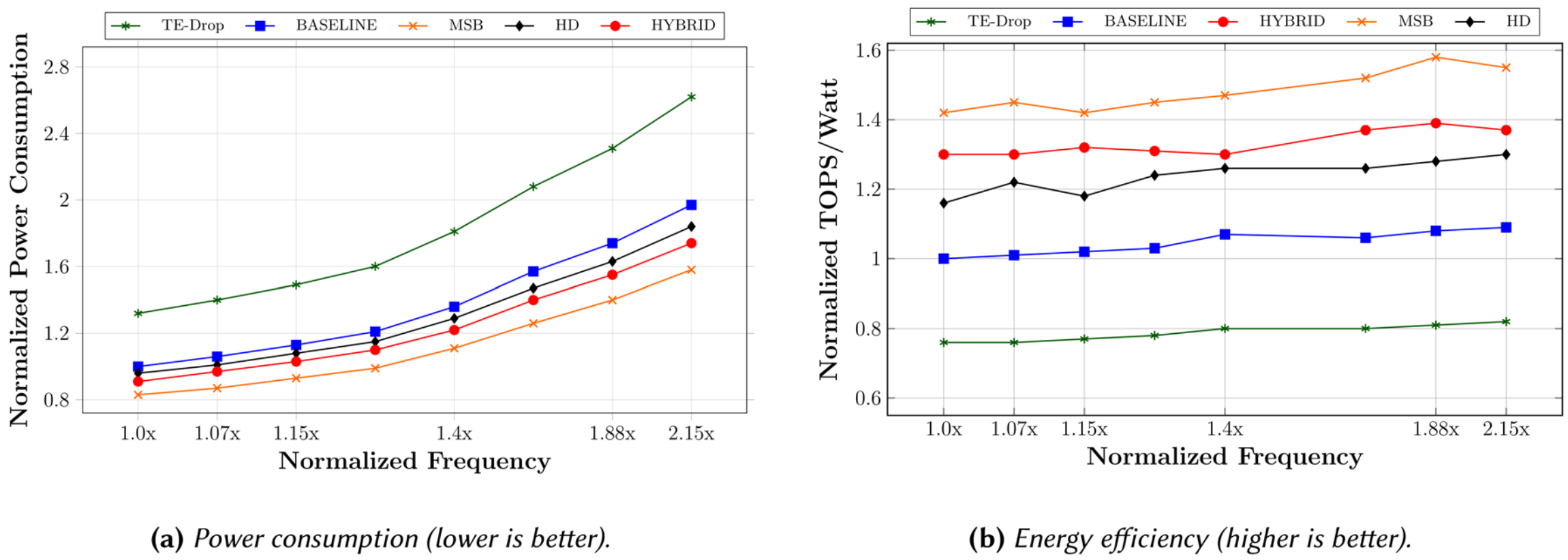

**(a)** *Power consumption (lower is better).* **(b)** *Energy efficiency (higher is better).*

**Fig. 8.** *Power and energy-efficiency trends across normalized frequency.*

(4) **MSB:** This scheme estimates timing criticality using operand magnitude. It counts the set bits in the upper half (MSBs) of the operand and maps the result to Fast/Mid/Slow tiers.

(5) **Hybrid:** This scheme combines the HD and MSB features into a single score and maps the result to Fast/Mid/Slow tiers.

### 5.2 Inference Accuracy

Figure 7 reports the normalized inference accuracy of the evaluated benchmarks under increasingly aggressive performance points. The benchmark suite spans diverse application domains, including vision benchmarks such as SVHN, GTSRB, CIFAR-10, MNIST, and FMNIST, text/NLP benchmarks such as Reuters and IMDB, and audio classification through AMNIST. This diversity already provides evidence that DSAC is not limited to a single dataset type or network structure. To further strengthen the evaluation for modern accelerator workloads, we also include ResNet-50 and BERT-Tiny. ResNet-50 represents a deeper convolutional benchmark, while BERT-Tiny represents an attention-based Transformer benchmark. For each benchmark, we first record an error-free reference accuracy at the baseline operating point and report all other accuracy values after normalizing to that reference.

Across all ten benchmarks, the normalized accuracies decline at different rates as frequency increases, illustrating the varying sensitivity of each workload to data-dependent timing errors. We observe a period of high stability in all schemes up to approximately 1.4× the baseline frequency, where the available timing slack is sufficient to absorb most variations. Beyond this point, accuracy begins to decline as the number of timing violations increases significantly.

The unprotected Baseline TPU exhibits the most rapid degradation across the evaluated workloads. In the GTSRB, CIFAR-10, and MNIST benchmarks, the Baseline accuracy collapses to 30% or lower at the extreme 2.15× frequency scaling point. While the reactive TE-DROP scheme provides some resilience by bypassing erroneous results, its accuracy also declines significantly as the clock period tightens. This degradation is particularly severe in MNIST, where TE-DROP accuracy falls to approximately 20% at 1.88× the normalized frequency, and in FMNIST and AMNIST, where it drops toward 45% and 65%, respectively, at the 2.15× mark.

In contrast, our proposed predictors (HD, MSB, and Hybrid) provide significantly better resilience by proactively adapting the timing budget. Across most benchmarks, the MSB and HD schemes consistently maintain the highest stability, often preserving near 100% normalized accuracy across the entire frequency range. While the Hybrid scheme tracks these top-performing predictors closely, it exhibits a minor accuracy dip in SVHN and AMNIST at the 2.15× mark compared to the standalone MSB and HD configurations. The ResNet-50 and BERT-Tiny results show that DSAC remains effective beyond the smaller benchmarks. On BERT-Tiny, the DSAC predictors maintain near 100% normalized accuracy across the full frequency range, while the Baseline drops significantly at 2.15×. While in case of ResNet-50, the HD and MSB predictors remain close to the error-free reference accuracy even under high frequency scaling, whereas TE-Drop experiences a larger accuracy reduction. Nevertheless, the Hybrid predictor remains highly robust, matching the top-tier performance in six out of ten benchmarks while consistently outperforming the Baseline and TE-DROP across all evaluated architectures.

### 5.3 Energy Efficiency

Figure 8(a) shows the power consumption for all of the comparative schemes. The power consumption for each scheme is normalized to the power consumed by the Baseline-TPU at the 1.0× baseline frequency. With the increasing operational frequency, power consumption steadily increases for all the schemes. However, our proposed predictors (HD, MSB, and Hybrid) exhibit lower power consumption compared to both TE-Drop and the Baseline-TPU at higher frequencies. Notably, beyond the 1.45× frequency mark, our proposed predictors consume less power than the unprotected Baseline because our proactive timing adaptation prevents the propagation of erroneous, high-toggle data that typically increases dynamic power throughout the systolic array.

The MSB predictor incurs the lowest power footprint among our designs, followed closely by the Hybrid predictor. In contrast, TE-Drop consistently exhibits the highest power consumption. The high overhead in TE-Drop results from its reactive error-detection mechanism, which requires energy-intensive cycles to stall the pipeline and manage timing violations. These results confirm that data-aware predictors allow for aggressive performance scaling with significantly lower power costs compared to traditional reactive mitigation techniques.

Figure 8(b) illustrates the normalized energy efficiency of each comparative scheme relative to that of the baseline at different normalized frequencies, measured in Tera-Operations Per Second per Watt (TOPS/Watt). Across the entire frequency spectrum, the energy efficiency generally follows an upward trend, indicating that the performance gains from frequency scaling outweigh the incremental power costs.

The reactive TE-Drop scheme remains the least efficient, consistently operating below the baseline with normalized TOPS/Watt values starting at 0.76 and only reaching approximately 0.82 at the maximum 2.15× frequency. This efficiency deficit highlights the significant power overhead of reactive stalls relative to the throughput they recover.

In contrast, our proposed predictors (HD, MSB, and Hybrid) significantly improve the energy-efficiency profile of the systolic array by proactively managing timing-critical cycles with minimal hardware overhead. Specifically, the MSB predictor emerges as the most energy-efficient scheme, delivering up to 1.55× better normalized TOPS/Watt compared to 1.1× baseline at maximum normalized frequency. This superior efficiency is primarily due to it having the lowest power consumption among all evaluated schemes, as the logic required for bit-counting in the operand's upper half is extremely lightweight. Our Hybrid and HD predictors follow closely, with the Hybrid predictor consistently providing approximately 1.3× better performance-per-watt than the baseline. Even at the highest performance points, these predictors maintain a substantial lead over traditional methods; for instance, at 2.15× frequency, the Hybrid and HD predictors are approximately 1.37× and 1.30× more efficient respectively, compared to the 0.82× recorded for the reactive TE-Drop scheme.

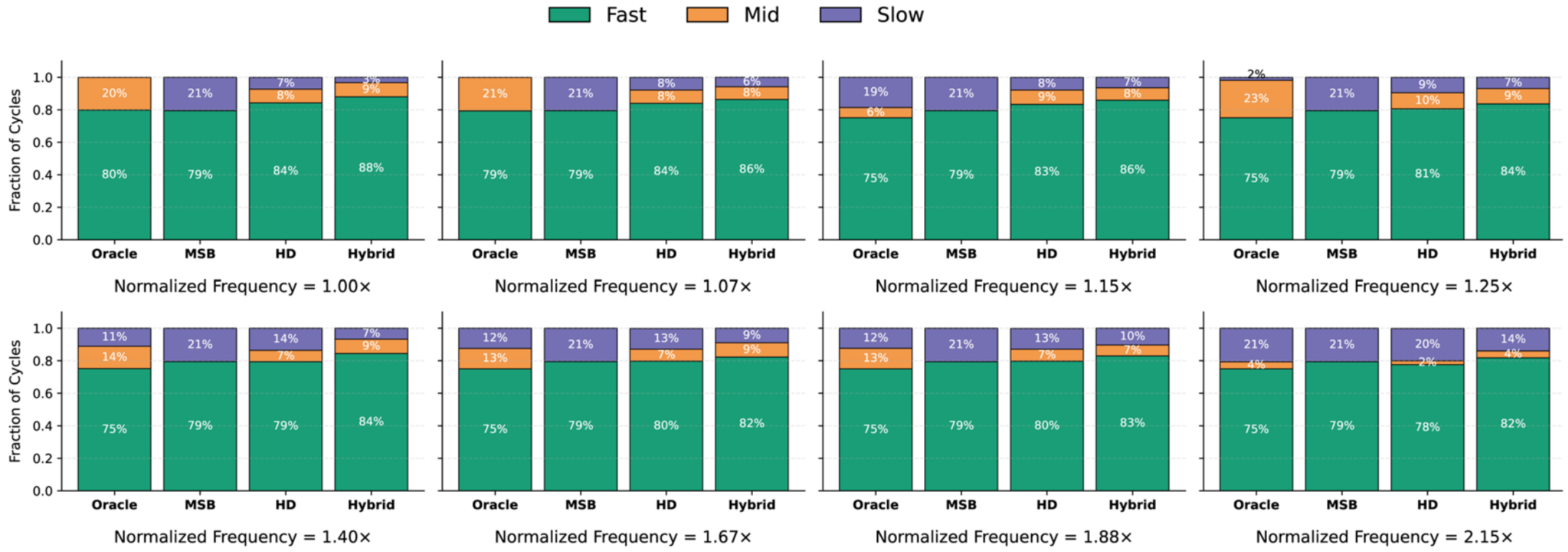

**Fig. 9.** *Timing-tier breakdown for the oracle and DSAC predictors. Averaged across the performance sweep, predictors successfully preserve a high fraction of Fast-tier execution, ensuring that systolic throughput remains dominated by the fast clock rather than rare worst-case delays.*

### 5.4 Predictor's Accuracy

Evaluation of the architectural effectiveness and operational reliability of DSAC involves studying both the timing-tier distribution and the predictor classification accuracy over a performance sweep from 1.0× to 2.15×. Fig. 9 reports the tier breakdown using three categories: Fast (zero hold cycles), Mid (1-cycle hold), and Slow (2-cycle holds). These results are reported at step granularity: each logical step is labeled by the worst tier requested among all active MACs, so the reported distribution directly reflects global throughput impact. In DSAC, tier labels map directly to dummy-hold overhead: Fast incurs 0 extra cycles, Mid incurs 1 extra cycle, and Slow incurs 2 extra cycles. We therefore quantify stall density as the average holds per step,

$$\text{holds/step} = \frac{N_{\text{Mid}} + 2N_{\text{Slow}}}{N_{\text{Total}}},$$

and the corresponding effective cycle count as $N_{\text{eff}}{=}N_{\text{Total}}{+}N_{\text{Mid}}{+}2N_{\text{Slow}}$, yielding a throughput penalty of $(N_{\text{eff}}/N_{\text{Total}}{-}1)$ relative to an all-Fast execution.

The timing-tier distribution reveals a direct correlation between clock frequency and threshold adaptation. At the 1.0× nominal frequency, the fast clock is sufficient for nearly all operations, including those with longer propagation delays. Consequently, the violation counter (viol_cnt) as shown in Algorithm 1, remains at zero, and the thresholds ($\tau_1$,$\tau_2$) maintain their initial settings. This leads to a distribution dominated by the Fast tier. As we move toward a more aggressive 2.15× normalized frequency, the shrinking clock period increases timing pressure, causing more operations to potentially violate the deadline. Once timing violations exceed the programmable limit $V_{\text{thr}}$, the closed-loop feedback logic proactively adjusts the tier thresholds. This adaptation is reflected in the varying sensitivity of the predictors. The MSB predictor shows remarkable consistency, maintaining nearly identical Fast/Mid/Slow ratios across all frequencies, indicating that its magnitude-based score partitions steps in a stable manner across operating points.

**Precision sensitivity of the MSB predictor.** To examine whether this behavior is tied to the INT8 `weight[7:4]` formulation, we also evaluated the MSB field extraction under INT4 and FP8-style configurations. For INT4, the predictor uses the upper two magnitude bits instead of the upper four INT8 bits. For FP8, the selected field is changed from integer upper-magnitude bits to the exponent field, which captures the numerical scale of the operand. Across these configurations, the MSB predictor shows nearly identical tier-distribution and application-level trends after threshold retuning. This indicates that DSAC does not depend on a fixed INT8 bit slice; rather, it depends on selecting a compact format-aware magnitude field that preserves the relative timing criticality of MAC operations. In the FP8 case, the correlation with multiplier path delay changes from integer upper-bit density to operand scale through the exponent field, with optional leading mantissa bits available when finer significand-level delay variation must be captured. The same Tier-LUT and feedback-controlled threshold mechanism therefore remain applicable across INT8, INT4, and FP8-style configurations.

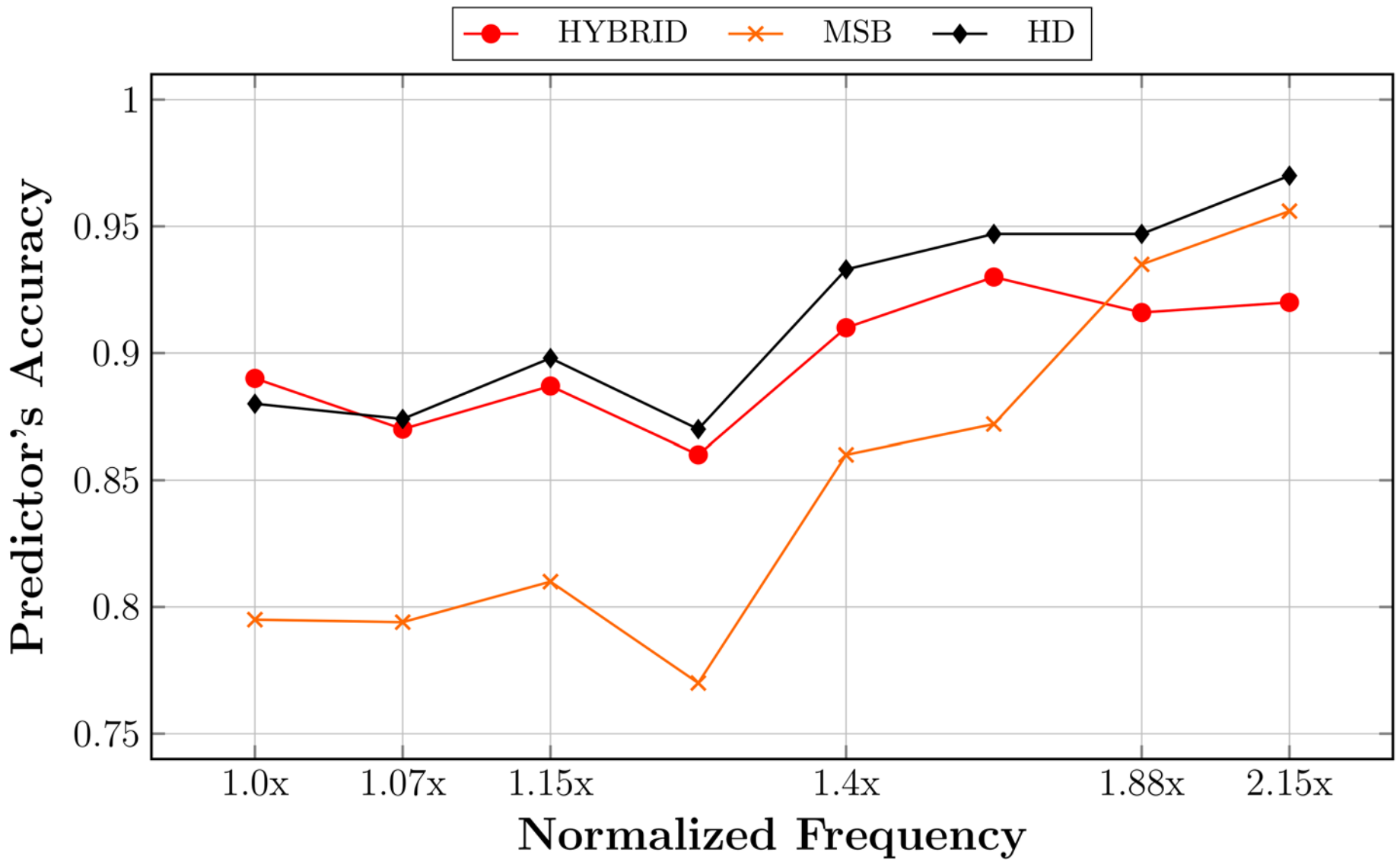

**Fig. 10.** *Predictor classification accuracy versus normalized frequency for Hybrid, MSB, and HD, measured as the fraction of cycles where the predicted tier (Fast/Mid/Slow) matches the oracle (Actual) tier.*

For the HD and Hybrid predictors, the fraction of Mid and Slow steps increases as frequency scales, indicating that the threshold logic allocates additional execution time as timing violations rise under the aggressive clock.

Even at the 2.15× scaling point, the measured stall density remains low, averaging 0.24–0.40 holds/step. Because the majority of operations (∼80%) continue to execute within a single fast-clock cycle, the net performance gain over the 1.0× baseline remains substantial. Fig. 10 further validates that these throughput gains are reliable and safe. High classification accuracy ensures two critical conditions: (i) performance preservation, by matching Fast oracle decisions and avoiding unnecessary holds, and (ii) timing safety, by correctly identifying Slow cases. The HD predictor reaches a peak accuracy of 0.97 at 2.15×, while Hybrid and MSB remain above 0.92 and 0.95, respectively. This high alignment confirms that lightweight operand-level features can effectively manage the multi-modal delay landscape of near-threshold TPUs without complex reactive recovery hardware.

To further validate that the predictor accuracy reported in this section is not merely a result of threshold tuning, we also evaluate the raw predictor scores before Fast/Mid/Slow tier classification. For each retained active MAC operation, we compare the raw predictor score with the STA-calibrated operation-level delay and measure the fraction of operation pairs for which the predictor-score ranking matches the delay ranking. This metric directly evaluates whether the operand-level feature preserves the relative timing criticality of MAC operations before any classification threshold is applied.

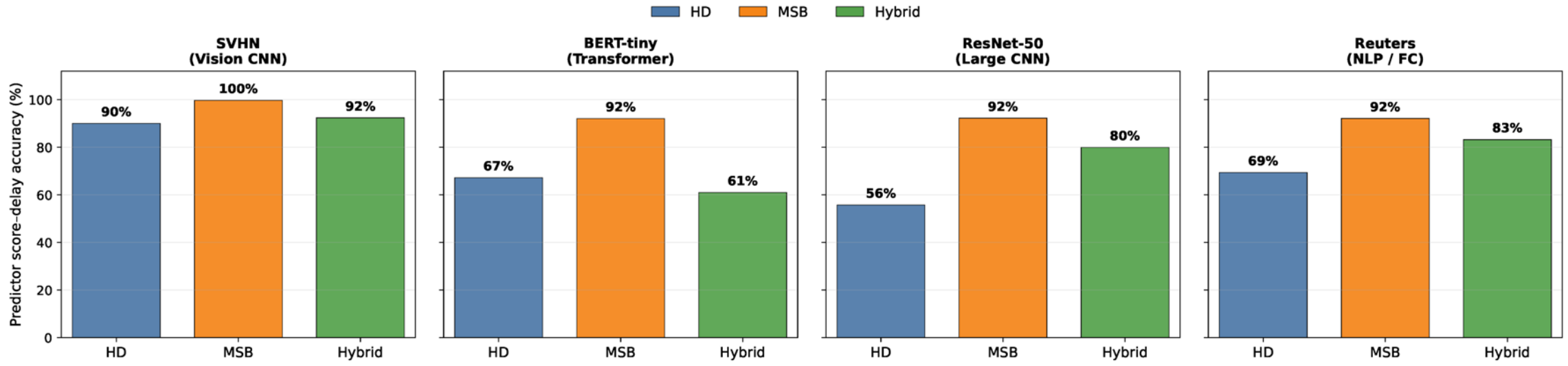

**Fig. 11.** *Cross-workload predictor–delay validation. The y-axis reports the percentage of active MAC-operation pairs for which the raw predictor-score ranking matches the STA-calibrated operation-delay ranking computed at the 1.0× normalized frequency. Higher values indicate stronger agreement between operand-level predictor scores and operation-level delay behavior before Fast/Mid/Slow thresholding is applied.*

Fig. 11 reports the predictor score–delay accuracy across four representative workload domains: SVHN for vision CNNs, BERT-tiny for Transformer-based inference, ResNet-50 for large CNNs, and Reuters for NLP/fully-connected inference. The MSB predictor provides the most stable behavior, achieving 100% accuracy on SVHN and 92% accuracy on BERT-tiny, ResNet-50, and Reuters. This result confirms that operand magnitude provides a consistent proxy for multiplier path delay across different network architectures and activation distributions. In contrast, the HD predictor is more workload dependent: it is highly accurate for SVHN at 90%, but drops to 67% on BERT-tiny and 56% on ResNet-50. This trend reflects the sensitivity of switching-based features to activation sparsity and input-transition patterns. The Hybrid predictor combines switching- and magnitude-related information and remains strong on SVHN, ResNet-50, and Reuters, achieving 92%, 80%, and 83%, respectively. On BERT-Tiny, however, Hybrid achieves 61%, which is lower than the standalone HD result of 67%. This does not contradict the benefit of feature fusion; rather, it shows that the fixed Hybrid weighting used in this work is not equally optimal for all workload types. In sparse Transformer activation streams, the HD term can become concentrated near very low toggle counts, reducing its discriminative value. Since the current Hybrid score gives higher weight to HD than to the MSB/magnitude term, this reduced HD variability can limit the benefit of fusion even when the magnitude feature remains informative.

These feature-level results provide a physical explanation for the decision-level predictor accuracy reported earlier. The high tier-classification accuracy is supported by raw predictor–delay alignment, showing that the predictors capture genuine operation-level timing behavior rather than merely fitting workload-specific thresholds. The workload dependence of HD is also consistent with the sparsity and input-transition behavior discussed in Section 5.6, while the stability of MSB explains why magnitude-aware prediction remains effective across workload domains. Overall, this analysis strengthens the methodological basis of DSAC by showing that the proposed operand-level predictors generalize across diverse DNN architectures and data patterns.

### 5.5 Throughput Impact of Dummy-Hold Cycles

The dummy-hold mechanism improves timing safety by extending the execution window of timing-critical MAC operations, but these extensions also introduce additional physical ticks. Therefore, the throughput impact of DSAC depends on how frequently operations are assigned to the Mid and Slow tiers. Fast-tier operations incur no extra delay, while

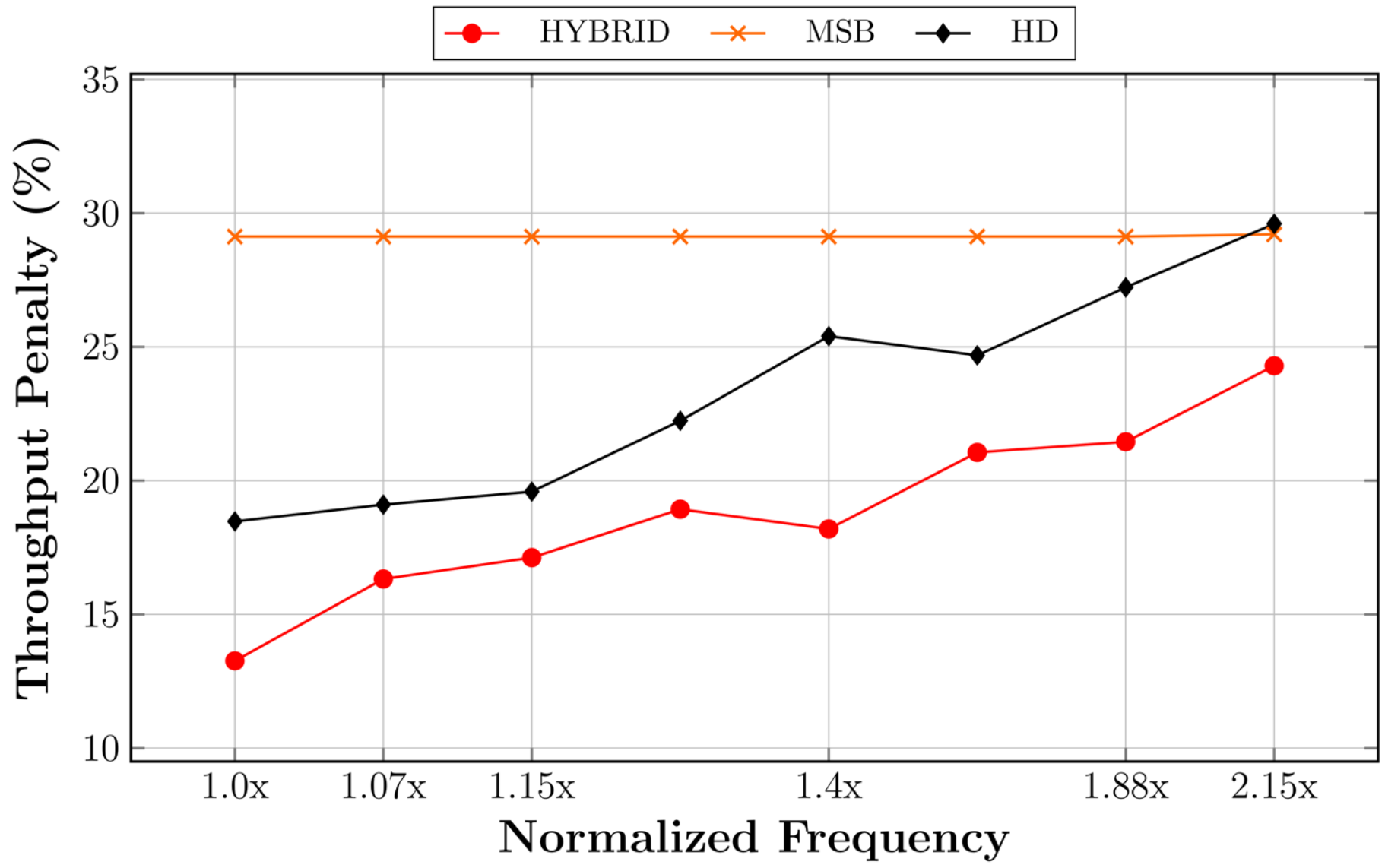

**Fig. 12.** *Throughput penalty versus normalized frequency for the Hybrid, MSB, and HD predictors. The penalty reflects the additional physical ticks introduced by Mid- and Slow-tier dummy-hold cycles during DSAC execution.*

Mid- and Slow-tier operations add one and two dummy-hold cycles, respectively. As a result, the effective execution time increases with the number of inserted hold cycles, even though the global fast reference clock remains unchanged.

Figure 12 reports the throughput penalty of the Hybrid, MSB, and HD predictors across the normalized frequency sweep. The penalty generally increases at more aggressive frequency-scaling points because the reduced clock period causes a larger fraction of operations to require Mid- or Slow-tier execution. This trend is most visible for the HD predictor, whose throughput penalty increases as frequency scales, indicating that more operations require additional timing margin. The Hybrid predictor maintains the lowest throughput penalty across most of the frequency range, showing that combining switching activity and operand magnitude helps avoid unnecessary hold-cycle insertion while still protecting timing-critical operations. The MSB predictor exhibits a comparatively stable but higher penalty, reflecting its more conservative tier assignment behavior.

Although dummy-hold insertion reduces raw throughput relative to an all-Fast execution, this overhead must be interpreted together with timing safety, accuracy, power, and energy efficiency. Without such timing extension, non-early-finishing MAC operations would be more likely to violate timing under aggressive near-threshold clock scaling, which can degrade inference accuracy. DSAC instead accepts a bounded throughput penalty for timing-critical operations in order to preserve application correctness. Moreover, the energy-efficiency results in Section 5.3 show that the throughput cost does not eliminate the benefit of DSAC: at the 2.15× frequency point, the MSB, Hybrid, and HD predictors still achieve improved normalized TOPS/Watt compared with the baseline and TE-Drop schemes. Thus, DSAC trades limited hold-cycle overhead for improved timing resilience and energy efficiency under aggressive near-threshold operation.

### 5.6 HD Predictor Behavior Under Sparse Activations

We further analyze the behavior of the HD predictor under sparse activation streams. The HD predictor estimates timing criticality from the number of bit toggles between consecutive activations. Therefore, when activation streams become highly sparse, such as after ReLU, many consecutive activation values may become zero or near-zero. This reduces the measured HD score and can shift many operations toward the Fast tier. As a result, an HD-only predictor can become overly optimistic under extreme sparsity, because low activation toggling does not always guarantee that the corresponding MAC operation is timing non-critical.

Figure 13 compares the HD-score distributions for dense pre-ReLU and sparse post-ReLU activation streams across CIFAR-10, Reuters, ResNet-50, and BERT-Tiny. The results show that post-ReLU sparsity significantly increases the fraction of low-HD and zero-HD operations. For Reuters, the zero-HD fraction increases from 10% in the dense stream to 81% in the sparse stream. Similarly, for ResNet-50, the zero-HD fraction increases from 10% to 82%. CIFAR-10 and BERT-Tiny also show strong zero-HD concentration, with sparse activation streams reaching 95% and 96% zero-HD operations, respectively.

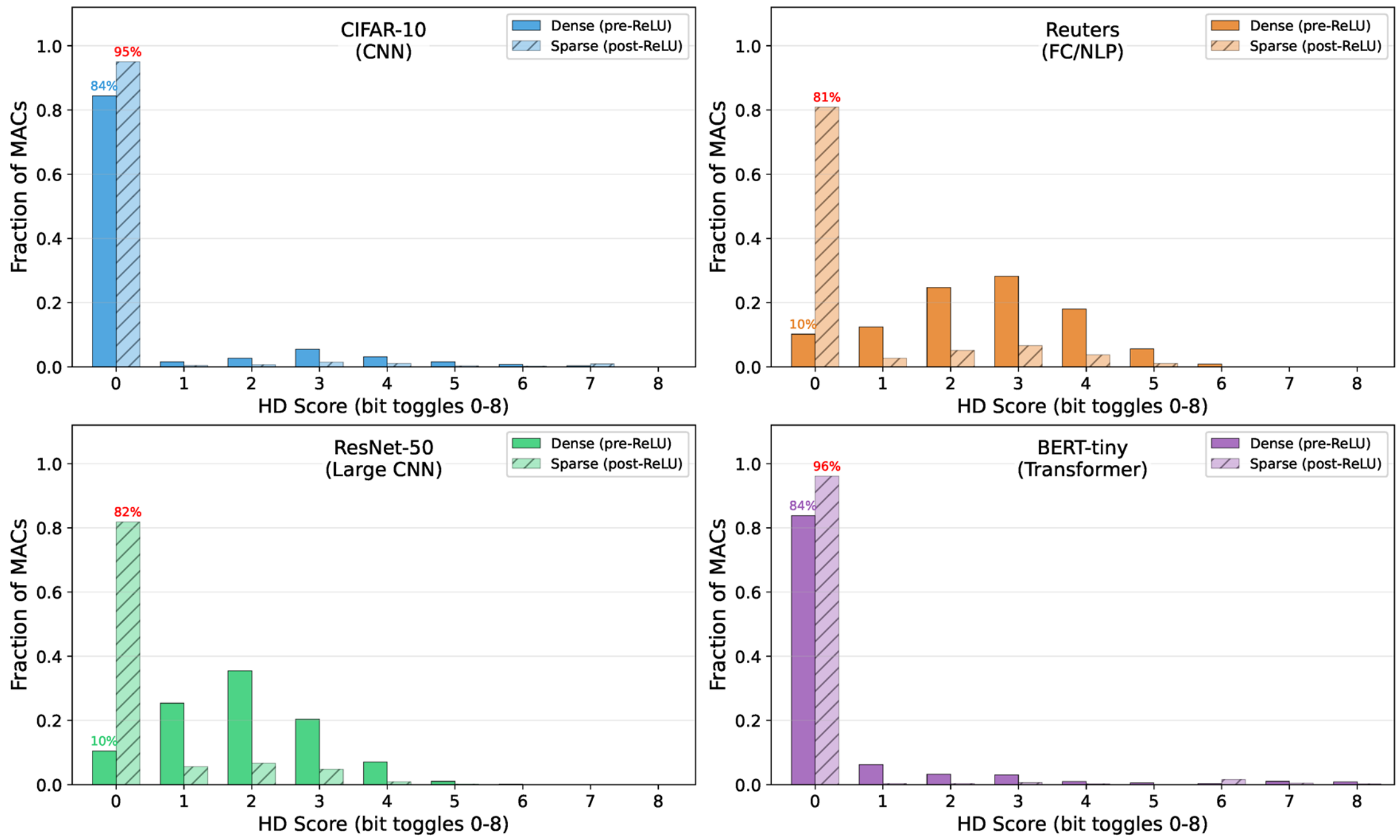

**Fig. 13.** *Effect of activation sparsity on HD-score distribution. Sparse post-ReLU activation streams shift a large fraction of operations toward low or zero HD scores, showing that HD-only prediction can become overly optimistic under zero-dominated activation patterns.*

This analysis confirms that HD alone may not be sufficient for all sparse activation patterns. However, DSAC is not restricted to HD-only prediction. The MSB predictor captures operand-magnitude information, while the Hybrid predictor combines activation switching and operand magnitude into a single criticality score. Therefore, even when sparse activations reduce the HD score, the MSB component can still identify operations with larger operand magnitudes and potentially longer logic paths. In addition, the closed-loop feedback controller monitors timing violations and adjusts tier thresholds at runtime, reducing the risk of persistent Fast-tier over-assignment under workload-specific sparse activation distributions.

| Scheme | Area Overhead (%) | Power Consumption | Energy Efficiency |
|---|---|---|---|
| Baseline | 0.0 | 1.95 | 1.09 |
| TE-Drop | 25.5 | 2.60 | 0.82 |
| MSB | 13.0 | 1.60 | 1.55 |
| HD | 25.4 | 1.90 | 1.30 |
| Hybrid | 25.1 | 1.80 | 1.37 |

**Table 2.** *Design overhead and performance benefits at normalized frequency 2.15×.*

### 5.7 Implementation Overhead

DSAC incurs area overhead primarily from the HD/MSB predictor logic and the local tier-control hardware. While these blocks are composed of simple bit-level operations (bit extraction, XOR, and lightweight popcount logic), the total overhead depends on the specific predictor instantiated and the inclusion of threshold-control circuitry.

Table 2 summarizes the hardware overheads and performance benefits of the comparative schemes at a normalized frequency of 2.15×. The MSB predictor remains the most compact design, incurring only a 13.0% area overhead relative to the baseline. Higher overheads are observed for the HD and Hybrid predictors (25.4% and 25.1%, respectively) because they incorporate additional bit-difference logic and threshold-control circuitry. Relative to the reactive TE-Drop scheme, which incurs a 25.5% area overhead, the MSB design reduces area by 10.0%, while the HD and Hybrid designs remain comparable in footprint.

At the 2.15× performance point, all DSAC predictors significantly reduce power consumption (1.60-1.90) compared to the reactive TE-Drop (2.6) and the unprotected Baseline (1.95). This proactive management translates to superior energy efficiency, with the MSB predictor reaching 1.55×, the Hybrid reaching 1.37×, and the HD reaching 1.30× better TOPS/W than the baseline. In contrast, the TE-Drop scheme's efficiency falls to 0.82. Overall, these results show that DSAC achieves proactive timing management and improved energy efficiency while maintaining a physical footprint that is competitive with, and in some cases smaller than, traditional reactive error-mitigation techniques.

### 5.8 Implementation Limitations and Practical Challenges

DSAC introduces fine-grained timing adaptation by combining lightweight predictor logic with local dummy-hold control. While this mechanism avoids global clock retuning and replay-based recovery, its practical implementation requires careful calibration and pipeline integration.

A key implementation consideration is the calibration of the tier-selection and feedback-control parameters. The thresholds ($\tau_1,\tau_2$) determine how predictor scores are mapped to Fast, Mid, and Slow timing tiers, while the feedback-controller parameters ($\Delta_1,\Delta_2,V_{thr}$) determine how quickly the system responds to observed timing violations. These parameters directly affect the trade-off between timing safety and throughput. Conservative thresholds reduce violations by assigning more operations to Mid or Slow tiers, but they also increase dummy-hold overhead. Aggressive thresholds preserve more Fast-tier execution, but they may increase the risk of timing violations under workload, voltage, or process variation. Although the closed-loop controller adapts the thresholds at runtime, the optimal controller configuration can still depend on the workload characteristics, operating voltage, process corner, and target timing-error tolerance. A more exhaustive sensitivity analysis of these controller parameters is therefore left for future work.

Another practical challenge is maintaining correct systolic dataflow alignment when dummy-hold cycles are inserted. In a weight-stationary array, activations advance horizontally while partial sums propagate vertically. When a Mid- or Slow-tier decision extends the execution window of a timing-critical operation, the corresponding activation and partial-sum state must be held consistently so that downstream processing elements observe temporally aligned operands. Improper hold control could lead to premature forwarding, duplicate operand consumption, or partial-sum misalignment between neighboring PEs. Therefore, the dummy-hold mechanism must be integrated with ready/valid gating or an equivalent local stall protocol that preserves wavefront synchronization while keeping the global fast reference clock unchanged.

These implementation considerations do not alter the DSAC mechanism itself, but they identify the main practical issues that must be addressed when deploying DSAC in a large systolic array: robust controller calibration and correct hold-cycle synchronization across the systolic dataflow.

## 6 Related Work

Prior work on near-threshold DNN accelerators largely falls into two categories: reactive schemes that detect timing errors and recover after they occur, and coarse-grained schemes that shift the operating point using voltage or frequency control. Razor represents the classic reactive approach, using double-sampling flip-flops to detect late signals and recover via replay or rollback, often within a global DVFS loop driven by the measured error rate [7]. ThunderVolt also targets aggressive near-threshold operation, but emphasizes voltage underscaling and policy-based recovery; when timing errors arise, recovery can include actions such as cycle stealing and dropped updates, with control applied at global, per-layer, or local scope [8]. SCISSORS takes an algorithmic path by embedding ABFT checks within the systolic fabric; errors are detected and corrected at block or array granularity, incurring checksum computation overhead rather than replay cycles [29]. EFFORT reduces dependence on DVFS by maintaining fixed-clock operation and using in-situ post-error correction through CostCo, coupled with opportunistic clock gating, to tolerate errors without replay [6]. GreenTPU moves further toward proactive control by learning error-prone activation sequences and selectively boosting local voltage rails, enabling prevention at row-level granularity through voltage-boost events [5]. Overall, most prior techniques remain reactive and input-agnostic, while GreenTPU is a key exception that introduces proactive intervention using data-dependent activation history.

A distinct alternative to clocked timing management is asynchronous circuit design, where request/acknowledge (REQ/ACK) handshaking replaces the global clock and each stage signals its own completion rather than committing to a fixed timing budget [30, 31]. This naturally accommodates data-dependent delay without requiring predictors or discrete timing tiers, but it introduces completion-detection circuitry and departs from the synchronous dataflow assumed by standard EDA flows and TPU-style systolic arrays. DSAC instead remains fully synchronous, using local dummy-hold control via the hold_req stall signal to enforce a pre-classified, known number of stall cycles $h$ for each MAC operation, rather than a bidirectional handshake negotiated between PEs. This avoids the completion-detection and negotiation overhead of a true asynchronous protocol while still resolving the same coordination concern raised by handshake-based designs—ensuring downstream PEs do not consume a result before it is ready.

Despite their effectiveness, existing techniques remain limited either by reactive recovery overheads, such as replay or dropped updates, or by coarse control scope, such as global, layer, block, or row decisions. Our work advances this trajectory by introducing proactive, operand-level prediction with per-PE, per-cycle control. Using lightweight Hamming distance and most-significant-bit features, each MAC estimates its cycle-level delay sensitivity and selects a timing tier (fast, mid, or slow) from a shared lookup table. The hybrid predictor combines both features to reduce unnecessary conservatism from HD-only decisions in dense activation regimes, while also reducing missed violations compared to MSB-only decisions. This enables fine-grained timing allocation under a fixed fast clock using local hold cycles, avoiding both voltage scaling and replay, and shifting near-threshold systolic arrays from recovery-centric operation toward prediction-driven timing orchestration.

| Technique | Proactive | Execution penalty | Control granularity | Data sensitivity | Clocking strategy |
|---|---|---|---|---|---|
| Razor [7] | ✗ | Replay rollback | Global | Agnostic | DVFS |
| ThunderVolt [8] | ✗ | Stolen cycle and dropped update | Global layer or local | Agnostic | Voltage underscaling |
| SCISSORS [29] | ✗ | In-array ABFT checksum overhead | Block array | Agnostic | DVFS |
| GreenTPU [5] | ✓ | Voltage boost events | Per row | Sequence based | Local voltage regulators |
| EFFORT [6] | ✗ | Same-cycle correction no replay | Local | Agnostic | Fixed |
| **DSAC** | ✓ | Hold cycles | Per PE per cycle | Operand based | Fixed fast clock |

**Table 3.** *Comparison of timing-error resilience techniques for near-threshold systolic accelerators. Proactive indicates whether mitigation is applied before an error occurs; execution penalty is the main runtime cost; control granularity is the scope of the decision or action; data sensitivity indicates whether decisions depend on input data; clocking strategy summarizes whether timing is managed via DVFS, voltage underscaling, local voltage regulators, or fixed-clock operation with local holds.*

## 7 Conclusion

Near-threshold operation can substantially reduce the energy consumption of TPU-style DNN accelerators, but it also increases delay sensitivity to process variation and operand-dependent switching activity. Under such conditions, conventional worst-case clocking leaves significant positive slack unused in common-case MAC operations, while aggressive clock scaling can increase timing violations and degrade application-level accuracy. This work addresses this problem by proposing Dynamic Slack-Aware Clocking (DSAC), a proactive timing adaptation framework that adjusts the effective execution window of MAC operations based on their predicted timing criticality.

DSAC uses lightweight operand-level predictors based on Hamming Distance (HD), Most-Significant-Bit (MSB) activity, and a Hybrid combination of both features to classify MAC operations into Fast, Mid, and Slow timing tiers. These tiers are enforced using local dummy-hold cycles under a fixed fast reference clock. As a result, timing-critical operations receive additional settling time without requiring global clock retuning, DVFS, or replay-based recovery. This makes DSAC suitable for TPU-style systolic arrays executing quantized DNN inference, where large numbers of MAC operations exhibit data-dependent timing variation while still requiring regular dataflow and high throughput.

Experimental results across ten quantized DNN benchmarks show that DSAC improves the energy efficiency of near-threshold TPU execution while preserving inference accuracy. At the aggressive 2.15× frequency-scaling point, the MSB predictor achieves the highest energy efficiency, reaching 1.55× normalized TOPS/Watt with only 13.0% area overhead. The Hybrid and HD predictors also improve energy efficiency, reaching 1.37× and 1.30× normalized TOPS/Watt, respectively, while maintaining competitive implementation cost. The proposed predictors achieve a peak tier-classification accuracy of 0.97 and keep average inference accuracy loss below 2%, showing that operand-level timing prediction can preserve application quality while improving performance per watt.

Overall, DSAC demonstrates that unused operand-dependent timing slack can be converted into energy-efficiency gains through lightweight prediction and local timing adaptation. Future work will study controller-parameter sensitivity, post-layout timing effects, and hold-cycle synchronization in larger systolic-array implementations.